\documentclass[12pt]{article}
\usepackage{amsfonts}
\usepackage{amsmath, amssymb}
\usepackage{youngtab}
\usepackage{hyperref}
\usepackage[dvips]{color}
\usepackage{psfrag}
\usepackage{feynmp}
\usepackage[pdftex]{graphicx}
\usepackage{braket}
\usepackage{bm}
\usepackage{subfigure}
\usepackage{subcaption}
\usepackage{tikz}

\usepackage{comment}
\includecomment{pdffig}

\allowdisplaybreaks[1]

\Yboxdim{5pt}

\newcommand{\sh}{{\mathrm{sh}}}
\newcommand{\ch}{{\mathrm{ch}}}

\makeatletter
    
    \@addtoreset{equation}{section}
  \makeatother

\usepackage{color}

\def\rem#1{}

\renewcommand{\title}[1]{\vbox{\center\LARGE{#1}}\vspace{5mm}}
\renewcommand{\author}[1]{\vbox{\center\large#1}\vspace{5mm}}

\begin{document}
\bibliographystyle{utphys}

\begin{titlepage}
\begin{center}
\vspace{5mm}
\hfill {\tt 
}\\
\vspace{20mm}

\title{
\LARGE
Quantized Coulomb branch  of 4d $\mathcal{N}=2$ $Sp(N)$ gauge theory and  spherical DAHA of $(C_N^{\vee}, C_N)$-type
}
\vspace{7mm}

Yutaka Yoshida

\vspace{6mm}

\vspace{3mm}
{\small {\it Department of Current Legal Study, Faculty of Law, Meiji Gakuin University, 1-2-37 
Shirokanedai, Minato-ku, Tokyo 108-8636, Japan}} \\
{\small {\it Institute for Mathematical Informatics, Meiji Gakuin University
1518 Kamikurata-cho, Totsuka-ku, Yokohama 244-8539, Japan}}
 \\

\end{center}

\vspace{7mm}
\abstract{
We study BPS loop operators in a 4d \(\mathcal{N}=2\) \(Sp(N)\) gauge theory with four hypermultiplets in the fundamental representation and one hypermultiplet in the antisymmetric representation. The algebra of BPS loop operators in the \(\Omega\)-background provides a deformation quantization of the Coulomb branch, which is expected to coincide with the quantized K-theoretic Coulomb branch in the mathematical literature.
For the rank-one case, i.e., \(Sp(1) \simeq SU(2)\), we show that the quantization of the Coulomb branch, evaluated using the supersymmetric localization formula, agrees with the polynomial representation of the spherical part of the double affine Hecke algebra (spherical DAHA) of \((C_1^{\vee}, C_1)\)-type.
For higher-rank cases, where \(N \geq 2\), we conjecture that the quantized Coulomb branch of the 4d \(\mathcal{N}=2\) \(Sp(N)\) gauge theory is isomorphic to the spherical DAHA of \((C_N^{\vee}, C_N)\)-type. As evidence for this conjecture, we show that the quantization of an 't Hooft loop agrees with the Koornwinder operator in the polynomial representation of the spherical DAHA.
}
\vfill

\end{titlepage}

\tableofcontents

\section{Introduction}

In three-dimensional (3d) \(\mathcal{N}=4\) supersymmetric (SUSY) gauge theories, there exists an interesting duality \cite{Intriligator:1996ex, Hanany:1996ie, deBoer:1996mp} known as 3d mirror symmetry. In this duality, the moduli space of Higgs vacua is isomorphic to the moduli space of Coulomb vacua of the dual theory. While the Higgs branch moduli space does not receive quantum corrections, the Coulomb branch moduli space receives both perturbative and non-perturbative corrections, making its analysis difficult.

A decade ago, a characterization of the Coulomb branch chiral ring (the coordinate ring of the Coulomb branch moduli space), consisting of vector multiplet scalars, bare monopole operators, and dressed monopole operators, was proposed in the mathematical literature \cite{Nakajima:2015txa, Braverman:2016wma} and in the physics literature \cite{Bullimore:2015lsa}. In the following, we refer to the Coulomb branch chiral ring simply as the Coulomb branch. It is known that interesting algebras, such as truncated shifted Yangians \cite{Braverman:2016pwk} and the rational Cherednik algebra \cite{Kodera:2016faj}, appear in the deformation quantization of the Coulomb branch, known as the {\it quantized Coulomb branch}. Here, the deformation quantization parameter corresponds to the \(\Omega\)-background parameter, and the product in the quantized Coulomb branch is identified with the operator product expansion in the presence of the \(\Omega\)-background \cite{Okuda:2019emk}; see also \cite{Assel:2019yzd}.

In the four-dimensional case, the vector multiplet scalars, bare monopole operators, and dressed monopole operators are lifted to Wilson loops, 't Hooft loops, and dyonic loops, respectively, in 4d \(\mathcal{N}=2\) supersymmetric gauge theories on \(S^1 \times \mathbb{R}^3\). In other words, 3d Coulomb branch operators arise from the Kaluza--Klein reduction along the \(S^1\) direction of 4d BPS loop operators. Consequently, the algebra of loop operators gives rise to the quantized Coulomb branch of 4d \(\mathcal{N}=2\) gauge theories on \(S^1 \times \mathbb{R}^3\), providing a trigonometric deformation of the 3d quantized Coulomb branch via Kaluza--Klein modes.
Furthermore, it is expected that the algebra of loop operators in certain gauge theories coincides with the quantized K-theoretic Coulomb branch. For example, in 4d \(\mathcal{N}=2^*\) \(U(N)\) gauge theory, one can explicitly show that the quantization of loop operators coincides with the polynomial representation of the spherical DAHA of \(\mathfrak{gl}_N\)-type \cite{Okuda:2019emk}, which is isomorphic to the quantized K-theoretic Coulomb branch.

A natural question arises as to whether different types of spherical DAHA appear in the Coulomb branch of 4d \(\mathcal{N}=2\) gauge theories or in the K-theoretic Coulomb branch. In this paper, we study the quantized Coulomb branch of 4d \(\mathcal{N}=2\) \(Sp(N)\) gauge theory with four fundamental hypermultiplets and one hypermultiplet in the antisymmetric representation. 
There is a structural reason why the spherical DAHA of \((C^{\vee}_N,C_N)\)-type is a natural candidate for the loop operator algebra in the present theory. Consider first the generic Cartan part of the algebra of Wilson and 't Hooft loops in the \(Sp(N)\) gauge theory on \(S^1\times\mathbb R^3\), before including the one-loop and monopole-bubbling contributions. The exponentiated electric and magnetic variables then parameterize two copies of \((\mathbb C^*)^N\). Therefore, the generic Cartan part of the classical loop operator algebra is identified with the algebra of Weyl-invariant Laurent polynomials in these variables:
\begin{align}
\mathbb{C}[(\mathbb{C}^{*})^N \times (\mathbb{C}^{*})^N]^{W_{Sp(N)}}\,,
\label{eq:classicalvev}
\end{align}
Here \(W_{Sp(N)}\) is the Weyl group of the gauge group \(Sp(N)\), i.e. the Weyl group of type \(C_N\).
On the DAHA side, the spherical DAHA of \((C^{\vee}_N,C_N)\)-type depends, in addition to \(q\), on the five parameters \(t_0,u_0,t_N,u_N,t\). When
 $ t_0=u_0=t_N=u_N=t=1,$
the spherical DAHA reduces to the \(W_{Sp(N)}\)-invariant quantum torus. Taking in addition the commutative limit \(q\to1\), it reduces precisely to the algebra in \eqref{eq:classicalvev}. It is also suggestive that the five parameters of the spherical DAHA can be matched with the five flavor fugacities associated with the four fundamental hypermultiplets and the antisymmetric hypermultiplet.

For the \(Sp(1) \simeq SU(2)\) gauge theory, we explicitly show that the deformation quantization of a generator of the  loop-operator algebra agrees with the polynomial representation of the spherical DAHA of \((C^{\vee}_1, C_1)\)-type.  For the \(Sp(N)\) gauge theory with \(N \geq 2\), we conjecture that the quantized Coulomb branch is isomorphic to the spherical DAHA of \((C^{\vee}_N, C_N)\)-type. We demonstrate that Wilson loops form a Laurent polynomial ring invariant under the Weyl group action of type \(C_N\), and that the quantization of the minimal charge 't Hooft loop coincides with the Koornwinder operator appearing in the polynomial representation of the spherical DAHA.  

This paper is organized as follows. In Section \ref{sec:section2}, we review the SUSY localization formula for BPS loop operators and its deformation quantization. The localization formula includes a contribution from the so-called monopole bubbling effect, which arises from the path integral over the moduli space of solutions to the Bogomol'nyi equation. In Section \ref{sec:monobub}, we review how monopole bubbling is evaluated in \(SU(2)\) gauge theory using branes in type IIB string theory. In Section \ref{sec:CCrank1}, we show that the algebra of loop operators in the \(Sp(1)\) gauge theory is isomorphic to the spherical DAHA of \((C^{\vee}_1, C_1)\)-type. In Section \ref{eq:CCNdaha}, we study the higher-rank case and show that the deformation quantization of loop operators agrees with elements of the polynomial representation of the spherical DAHA of \((C^{\vee}_N, C_N)\)-type. Finally, Section \ref{sec:summary} is devoted to summary and future directions.

\section{BPS loops in 4d \(\mathcal{N}=2\) gauge theories on \(S^1 \times \mathbb{R}^3\)}
\label{sec:section2}

\subsection{SUSY localization formula}
In this section, we explain the SUSY localization formula for BPS loop operators in 4d \(\mathcal{N}=2\) SUSY gauge theory on \(S^1 \times \mathbb{R}^3\) \cite{Ito:2011ea}. The vacuum expectation value (VEV) of a BPS loop operator \(L_{({\bm p},{\bm q})}\) is defined by the following supersymmetric index:
\begin{align}
\langle L_{({\bm p},{\bm q})} \rangle := \mathrm{Tr}_{\mathcal{H}_{({\bm p},{\bm q})}(\mathbb{R}^3)} (-1)^{\sf F} e^{\epsilon_+ ({\sf J}_3+{\sf R})} \prod_{f} e^{{\sf F}_f m_f}\,.
\label{eq:loopop}
\end{align}
Here, \(({ \bm p}, {\bm q}) \in \Lambda_{\rm cw}(G) \times \Lambda_{\rm w}(G)/W_G\), where \(\Lambda_{\rm w}(G)\) (resp. \(\Lambda_{\rm cw}(G)\)) denotes the weight (resp. coweight) lattice of the Lie algebra of the gauge group \(G\). \(W_G\) is the Weyl group of \(G\). 
In this article, we refer to \({\bm p}\) (resp. \({\bm q}\)) as the magnetic (resp. electric) charge. \(\mathcal{H}_{({\bm p},{\bm q})}(\mathbb{R}^3)\) is the Hilbert space in the presence of the loop operator.
\({\sf F}\) is the fermion number operator. \({\sf J}_3\) generates spacetime rotations in \(\mathbb{R}^2_{\epsilon_+} := \mathbb{R}^2 \subset \mathbb{R}^3\). \({\sf R}\) generates \(U(1) \subset SU(2)_H\), where \(SU(2)_H\) is the R-symmetry group of the 4d \(\mathcal{N}=2\) theory. \({\sf F}_f\) denote the generators of the maximal torus of the flavor symmetry group acting on the hypermultiplets.
The parameters \(\epsilon_+\) and \(m_f\) are the fugacities associated with these generators. \(\epsilon_+\) is called the \(\Omega\)-background parameter, and \(m_f\) is called the flavor fugacity.

The BPS condition constrains the positions of loop operators in the spacetime \(S^1 \times \mathbb{R}^3\). When \(\epsilon_+ \neq 0\), a loop operator wraps \(S^1\) and is located at \((0,0,x^3) \in \mathbb{R}^2_{\epsilon_+} \times \mathbb{R} = \mathbb{R}^3\), where \(x^3\) is arbitrary.
The correlation functions of loop operators may depend on their ordering. When \(\epsilon_+ = 0\), the VEVs of loop operators are independent of the insertion points in \(\mathbb{R}^3\). 

In the path integral formalism, the VEV of $L_{({\bm p},{\bm q})}$ is defined as the VEV of a Wilson loop with electric charge ${\bm q}$ in the monopole background with magnetic charge ${\bm p}$ \cite{Kapustin:2006pk}, and can be evaluated using SUSY localization \cite{Ito:2011ea}. To define magnetically charged loop operators (\({\bm p} \neq 0\)), we impose a singular boundary condition near the origin of \(\mathbb{R}^3\). Let \((A_{\mu}, \sigma, \varphi)\) be the gauge field and two real adjoint scalars in the 4d \(\mathcal{N}=2\) vector multiplet.
The BPS (singular) boundary condition for the gauge field \(A_{i=1,2,3}\) and the real scalar \(\sigma\) in the vector multiplet is given by:
\begin{align}
A & = \frac{\bm p}{2} \cos \theta d\theta + \cdots \,, \,\, \sigma =\frac{\bm p}{2r}+\cdots \,.
\end{align}
Here, \((r, \theta, \phi)\) are the polar coordinates of \(\mathbb{R}^3\) centered at the loop operator. These boundary conditions satisfy the Bogomol'nyi equation:
$F_A = *_{3} D_A \sigma$.

The SUSY localization formula is given by:
\begin{align}
\langle L_{({\bm p},{\bm  q})} \rangle &= 
 \sum_{w \in W_G}   e^{ w( {\bm p}) \cdot {\bm b} +w({\bm q}) \cdot {\bm a} }
Z_{1\text{-loop}} (w ({\bm p}), {\bm a} , {\bm m}, \epsilon_+) 
\nonumber \\
&\quad + \sum_{ \tilde{\bm p} \in {\bm p}+\Lambda_{\rm cr}(G) \atop \|\tilde{\bm p}\| < \|{\bm p} \|} e^{ \tilde{\bm p} \cdot {\bm b}    }
Z_{1\text{-loop}} (\tilde{\bm p}, {\bm a} , {\bm m}, \epsilon_+) Z_{\text{mono}} ({\bm p}, \tilde{\bm p}, {\bm q}, {\bm a},{\bm m}, \epsilon_+)\,. 
\label{eq:localization}
\end{align}
Here, \(\Lambda_{\rm cr}(G)\) is the coroot lattice, \(\|\bm p\|\) denotes the norm of \(\bm p\), \(\alpha \cdot {\bm p}\) represents the inner product of \(\alpha\) and \(\bm p\), and \(w({\bm p})\) denotes the Weyl group action on \(\bm p\).
The one-loop determinant \(Z_{1\text{-loop}}\) consists of the one-loop determinant \(Z^{\text{v.m.}}_{1\text{-loop}}\) of the vector multiplet and the one-loop determinant \(Z^{\text{h.m.}}_{1\text{-loop}}\) of a hypermultiplet in a representation \(R\)\footnote{In this paper, we refer to a hypermultiplet belonging to a representation \(R \oplus R^*\) of the gauge group simply as a hypermultiplet in \(R\).}:
\begin{align}
Z_{1\text{-loop}} ({\bm p}, {\bm a} , {\bm m}, \epsilon_+) 
=Z^{\text{v.m.}}_{1\text{-loop}} ({\bm p}, {\bm a} , \epsilon_+)Z^{\text{h.m.}}_{1\text{-loop}}
({\bm p}, {\bm a} , {\bm m}, \epsilon_+),
\label{eq:oneloopG}
\end{align}
with  
\begin{align}
Z^{\text{v.m.}}_{1\mathchar `-\text{loop}}({\bm a},\bm{p}, \epsilon_+)= 
 \left[ \prod_{\alpha:\text{root} } \prod_{k=0}^{|\alpha \cdot \bm{p}|-1} \sh  \Bigl ( \alpha \cdot {\bm a} - ( |\alpha \cdot \bm{p}| - 2k ) \epsilon_+ \Bigr) \right] ^{-\frac{1}{2}},
\label{eq:1loopvec}
\end{align}
\begin{align}
Z^{\text{h.m.}}_{1 \mathchar `-\text{loop}}({\bm a}, {\bm m}, \bm{p}; \epsilon_+)=  
   \left[ \prod_{{\rm w} \in wt(R)} \prod_{\mu \in wt(R_F)} \prod_{k=0}^{|{\rm w} \cdot \bm{p}|-1}{\rm sh} \Bigl ( {\rm w} \cdot {\bm a} +\mu \cdot {\bm m} - (|{\rm w} \cdot \bm{p}| -1-2k ) \epsilon_+ \Bigr) \right]^{\frac{1}{2}}\,.
\label{eq:1loophy}
\end{align}
Here, \(\sh(x) := 2 \sinh(x/2)\).
${\bm a} :=  (a_1, a_2, \dots, a_{{\rm rank}(G)})$ is a holomorphic combination of the gauge field $A_0$ and the vector multiplet scalar $\varphi$.  
${\bm b} := (b_1, b_2, \dots, b_{{\rm rank}(G)})$ is a holomorphic combination of the magnetic charge fugacity and the vector multiplet scalar $\sigma$.  
${\bm a}$ and ${\bm b}$ take values in the complexification of the Cartan subalgebra of the Lie algebra of $G$, which are fixed by the boundary conditions of  
$A_0$, $\sigma$, and $\varphi$ at spatial infinity.  
$wt(R)$ (resp. $wt(R_F)$) denotes the weights of a representation $R$ (resp. $R_F$) of the gauge group (resp. flavor symmetry group).  
${\bm m} := (m_1, m_2, \dots, m_{{\rm rank}(G_F)})$ represents the flavor fugacities of the hypermultiplet, which take values in the Cartan subalgebra of  
the Lie algebra of the flavor symmetry group $G_F$.  
In Section \ref{sec:monobub}, we explain how to evaluate the monopole bubbling effect $Z_{\rm mono}$ in the localization formula.

\subsection{Deformation quantization and algebra of loop operators}
\label{sec:defqu}

Following \cite{Ito:2011ea}, we define the deformation quantization of the VEVs of loop operators using the Weyl--Wigner transformation, also known as Weyl quantization.
The Weyl--Wigner transformation \(\widehat{f}(\widehat{\bm a}, \widehat{\bm b})\) of a function \(f({\bm a}, {\bm b})\) can be computed efficiently using the formula:
\begin{align}
\widehat{f}(\widehat{\bm a}, \widehat{\bm b})= \exp \left(-\epsilon_+ \sum_{i=1}^{{\rm rank}(G)} \partial_{a_i} \partial_{b_i} \right) {f}({\bm a}, {\bm b}) \Big|_{{\bm a} \mapsto \hat{\bm a}, {\bm b} \mapsto \hat{\bm b}}\,,
\label{eq:weylwig}
\end{align}
where \(\hat{a}_i\) and \(\hat{b}_i\) satisfy the commutation relations
\begin{align}
[\hat{b}_i, \hat{a}_j] = -2 \epsilon_+ \delta_{ij}, \quad [\hat{a}_i, \hat{a}_j] = 0, \quad [\hat{b}_i, \hat{b}_j] = 0\,.
\label{eq:comm}
\end{align}
On the right-hand side of \eqref{eq:weylwig}, we assume that the operators \(\hat{a}_i\) are placed to the left of \(\hat{b}_i\).

As discussed in \cite{Ito:2011ea}, correlation functions satisfy the following relation:
\begin{align}
\langle L_{({\bm p}_1, {\bm q}_1)}(x^3_1) L_{({\bm p}_2, {\bm q}_2)}(x^3_2) \cdots L_{({\bm p}_n, {\bm q}_n)}(x^3_n)\rangle=
\langle L_{({\bm p}_1, {\bm q}_1)} \rangle * \langle  L_{({\bm p}_2, {\bm q}_2)} \rangle * \cdots * \langle L_{({\bm p}_n, {\bm q}_n)}\rangle.
\label{eq:corMoyal}
\end{align}
Here, \(x^3_i\) denotes the \(x^3\)-coordinate of \(L_{({\bm p}_i, {\bm q}_i)}\) and is assumed to satisfy \(x^3_i > x^3_{i+1}\) for \(i=1,\dots, n-1\).
The symbol \(*\) represents the Moyal product, defined by
\begin{align}
{f * g } ({\bm a}, {\bm b}) := \exp \Bigl[ \epsilon_+  \sum_{i=1}^{\mathrm{rank}(G)}(\partial_{a_i} \partial_{b^{\prime}_i} - \partial_{a^{\prime}_i} \partial_{b_i} ) \Bigr]
 f({\bm a}, {\bm b})  g({\bm a}^{\prime}, {\bm b}^{\prime}) \Bigr|_{{\bm a}^{\prime} \mapsto {\bm a} \atop {\bm b}^{\prime} \mapsto {\bm b}}  
\end{align}
and satisfies the relation
\begin{align}
\widehat{f * g }( \widehat{\bm a}, \widehat{\bm b}) = \hat{f}(\widehat{\bm a}, \widehat{\bm b}) \, \hat{g}(\widehat{\bm a}, \widehat{\bm b}) \,.
\end{align}

Applying the deformation quantization to \eqref{eq:corMoyal}, we obtain
\begin{align}
&\text{The Weyl--Wigner transformation of } \langle L_{({\bm p}_1, {\bm q}_1)} L_{({\bm p}_2, {\bm q}_2)} \cdots L_{({\bm p}_n, {\bm q}_n)}\rangle \nonumber \\
&=
\hat{L}_{({\bm p}_1, {\bm q}_1)} \hat{L}_{({\bm p}_2, {\bm q}_2)} \cdots \hat{L}_{({\bm p}_n, {\bm q}_n)}.
\label{eq:defcol}
\end{align}
Here, we define \(\hat{L}_{({\bm p},{\bm q})}:=\widehat{\langle L_{({\bm p},{\bm q})} \rangle }\).
Thus, the deformation quantization of the VEVs of loop operators is identified with the operator product of loop operators \cite{Okuda:2019emk}, and the algebra of loop operators is defined via the operator product expansion of loop operators. For 3d \(\mathcal{N}=4\) gauge theories, the same procedure \eqref{eq:weylwig}--\eqref{eq:defcol} defines the algebra of Coulomb branch operators, i.e., the algebra of Coulomb branch scalars and monopole operators. It was shown in \cite{Okuda:2019emk} that the algebra of Coulomb branch operators defined using the Moyal product and the Weyl--Wigner transformation of the localization formula agrees with the (abelianized) quantized Coulomb branch in the sense of \cite{Bullimore:2015lsa}.

Loop operators \(L_{({\bm p},{\bm 0} )}\) and \(L_{({\bm 0},{\bm q} )}\), with \({\bm 0} := (0,0,\dots,0)\), are referred to as the BPS 't Hooft loop and the BPS Wilson loop, respectively. The loop operator \(L_{({\bm p},{\bm q} )}\) with \({\bm q} \neq {\bm 0}\) and \({\bm p} \neq {\bm 0}\) is called a BPS dyonic loop (also known as a Wilson--'t Hooft loop). Since the one-loop determinant \eqref{eq:oneloopG} becomes trivial for zero magnetic charge \({\bm p}=0\), the VEV of the Wilson loop is simply given by the character of a representation of \(G\) labeled by the highest weight \({\bm q}\):
\begin{align}
\langle L_{({\bm 0},{\bm  q} ) } \rangle&=
 \sum_{w \in W_G}   e^{ w({\bm q}) \cdot {\bm a} }.
\end{align}
For example, if we consider \(G=U(N)\) and \({\bm q}=(1,0, \dots, 0)\), the VEV of the Wilson loop is given by
\begin{align}
\langle L_{({\bm 0},{\bm  q} ) } \rangle&=\sum_{i=1}^{N} e^{a_i}.
\end{align}
In general, defining \(x_i:=e^{-a_i}\), the VEV of a Wilson loop belongs to the \(W_G\)-invariant Laurent polynomial ring:
\begin{align}
\langle L_{({\bm 0},{\bm  q} ) } \rangle \in \mathbb{C} [x^{\pm1}_1, x^{\pm1}_2, \dots, x^{\pm1}_{{\rm rank}(G)}]^{W_{G}}\,,
\end{align}
and the algebra of BPS Wilson loops is identified with this \(W_G\)-invariant Laurent polynomial ring. From the commutation relation \eqref{eq:comm}, the quantization of 't Hooft loops and dyonic loops can be regarded as difference operators acting on the algebra of Wilson loops, i.e., the \(W_G\)-invariant Laurent polynomial ring.

\section{Monopole bubbling effect}
\label{sec:monobub}

The monopole bubbling effect \(Z_{\rm mono}\) in \eqref{eq:localization} arises from the path integral over the moduli space of solutions to the Bogomol'nyi equation with a reduced magnetic charge \(\tilde{\bm p}\), measured at infinity in \(\mathbb{R}^3\), where
\(\tilde{\bm p} \in {\bm p}+\Lambda_{\rm cr}(G)\) and \(\|\tilde{\bm p}\| < \|{\bm p} \|\).
For \(G=SU(N)\), the monopole bubbling effect in 't Hooft loops was originally evaluated in \cite{Gomis:2011pf, Ito:2011ea} using Kronheimer's correspondence \cite{Kronheimer:MTh}. Kronheimer's correspondence implies that the moduli space of solutions to the Bogomol'nyi equation with a reduced magnetic charge is a subset of the instanton moduli space on a Taub--NUT space. Consequently, \(Z_{\rm mono}\) is expected to be obtained by a certain truncation of Nekrasov's formula for the 5d (K-theoretic) instanton partition function. However, it was found that the \(Z_{\rm mono}\) obtained via Kronheimer's correspondence agrees only partially with the Verlinde loop operators in Liouville CFT, which should correspond to BPS loop operators in the AGT dictionary \cite{Drukker:2009tz, Drukker:2009id, Alday:2009fs}.

This discrepancy was resolved in \cite{Brennan:2018rcn}, where a D-brane construction for the moduli space of monopole bubbling was proposed.
In this approach, \(Z_{\rm mono}\) is given by the Witten index of a quiver supersymmetric quantum mechanics (SQM) associated with the low-energy worldvolume theory on D1-branes. If the FI parameter \(\zeta\) of the SQM is at a generic point in FI-parameter space, the Witten index is computed using the Jeffrey--Kirwan (JK) residue \cite{Hori:2014tda, Hwang:2014uwa} and coincides with the truncated Nekrasov partition function.
On the other hand, when the FI parameter is zero, additional  states in the SQM may contribute to the Witten index. In fact, the D-brane configuration for \(SU(N)\) gauge theory suggests that the FI parameter is zero. Interestingly,
computations in various examples suggest that the genuine monopole bubbling contribution takes the following form:
\begin{align}
Z_{\rm mono}=Z^{(\zeta)}_{\rm JK}+Z^{(\zeta)}_{\rm extra}\,.
\end{align}
Here, \(Z^{(\zeta)}_{\rm JK}\) is the Witten index for monopole bubbling, evaluated at a generic point \(\zeta\) in FI-parameter space. \(Z^{(\zeta)}_{\rm JK}\) is computed using the JK residue of the quiver SQM or an instanton partition function. In this paper, we refer to \(Z^{(\zeta)}_{\rm JK}\) as the {\it JK part}.
\(Z^{(\zeta)}_{\rm extra}\) represents the additional contribution in the SQM.

The value of \(Z^{(\zeta)}_{\rm JK}\) may change discontinuously when the FI parameter \(\zeta\) crosses a codimension-one wall in FI-parameter space. This phenomenon is known as {\it wall-crossing}. Note that the JK part is an equivariant index of the moduli space of Higgs branch vacua in the SQM. Thus, the wall-crossing phenomenon in SQM is the same as that in the mathematical literature, where the index depends on the stability condition.
\(Z^{(\zeta)}_{\rm extra}\) also depends on the FI parameter. However,
the sum \(Z^{(\zeta)}_{\rm JK} + Z^{(\zeta)}_{\rm extra}\) is independent of the choice of FI parameter and gives the Witten index at zero FI parameter: \(Z_{\rm mono}\).

In \cite{Brennan:2018rcn}, \(Z^{(\zeta)}_{\rm extra}\) for 't Hooft loops in \(SU(N)\) gauge theory was evaluated using the Born--Oppenheimer approximation, which is complicated even for minimal magnetic charge and becomes increasingly complicated for higher magnetic charges \({\bm p}\). Besides this approach,
there are two alternative methods to compute \(Z^{(\zeta)}_{\rm extra}\).
One is to use the complete brane setup explained in Section \ref{sec:imp},
and the other is to subtract decoupled states from \(Z^{(\zeta)}_{\rm JK}\), as explained in Section \ref{sec:subt}.

\subsection{Naive brane setup for monopole bubbling in \(U(N)\) and \(SU(N)\) gauge theories}

\begin{table}[t]
\begin{center}
\begin{tabular}{c|cccc|cc|cccc}
&0&1&2&3&4&5&6&7&8&9
\\
\hline
D3&$\times$&$\times$&$\times$&$\times$& && 
\\
D7&$\times$&$\times$&$\times$&$\times$&&&$\times$&$\times$&$\times$&$\times$
\\
\hline
NS5 &$\times$&&&&&$\times$&$\times$&$\times$&$\times$&$\times$\\
D1 &$\times$&&&&$\times$&&&&&
\\
\hline
D5 &$\times$&&&&$\times$&&$\times$&$\times$&$\times$&$\times$
\end{tabular}
\end{center}
\caption{The brane configuration for the 't Hooft loop and the monopole bubbling effect in 4d $\mathcal{N}=2$ gauge theory.
 The symbol $\times$ represents the directions in which branes extend.}
\label{table:brane}
\end{table}

\begin{pdffig}
\begin{figure}[thb]
\centering
\subfigure[]{\label{fig:subbrane1}
\includegraphics[height=4cm]{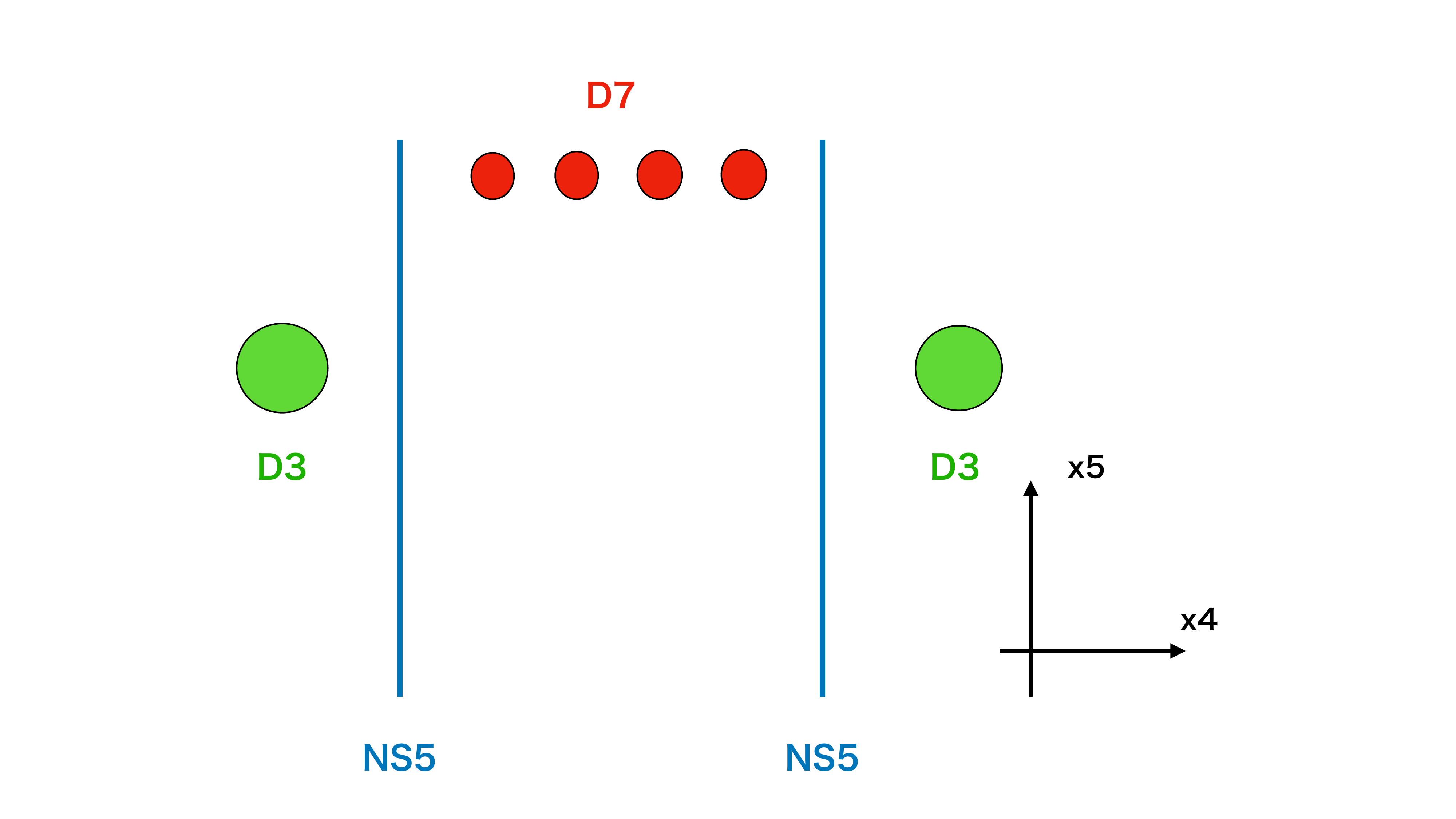}
}
\subfigure[]{\label{fig:subbrane2}
\includegraphics[height=4cm]{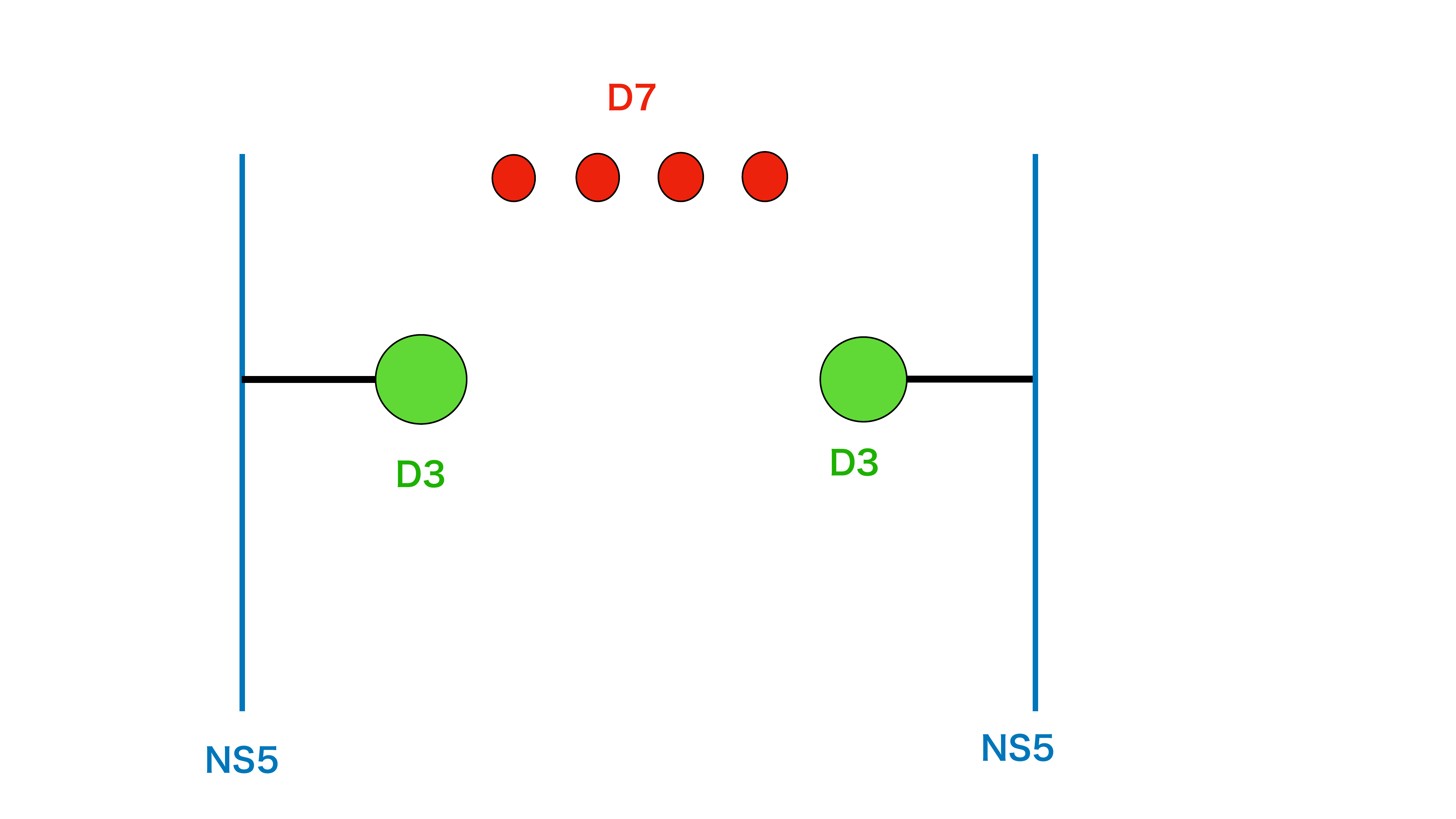}}
\caption{(a): A brane configuration in the \((x^4,x^5)\)-plane for an 't Hooft loop with magnetic charge \({\bm p}=(1,-1)\) (resp. \({\bm p}=1\)) in \(U(2)\) (resp. \(SU(2)\)) gauge theory with four hypermultiplets. 
The red and green circles represent a D7-brane and a D3-brane, respectively. The blue line represents an NS5-brane. (b): Another brane configuration corresponding to an 't Hooft loop. Figures (a) and (b) are related by the Hanany--Witten effect: 
when a D3-brane crosses an NS5-brane, a D1-brane (denoted by a black line) is either created or annihilated.}
\label{fig:brane1}
\end{figure}
\end{pdffig}


\begin{pdffig}
\begin{figure}[thb]
\centering
\subfigure[]{\label{fig:subbrane3}
\includegraphics[height=4cm]{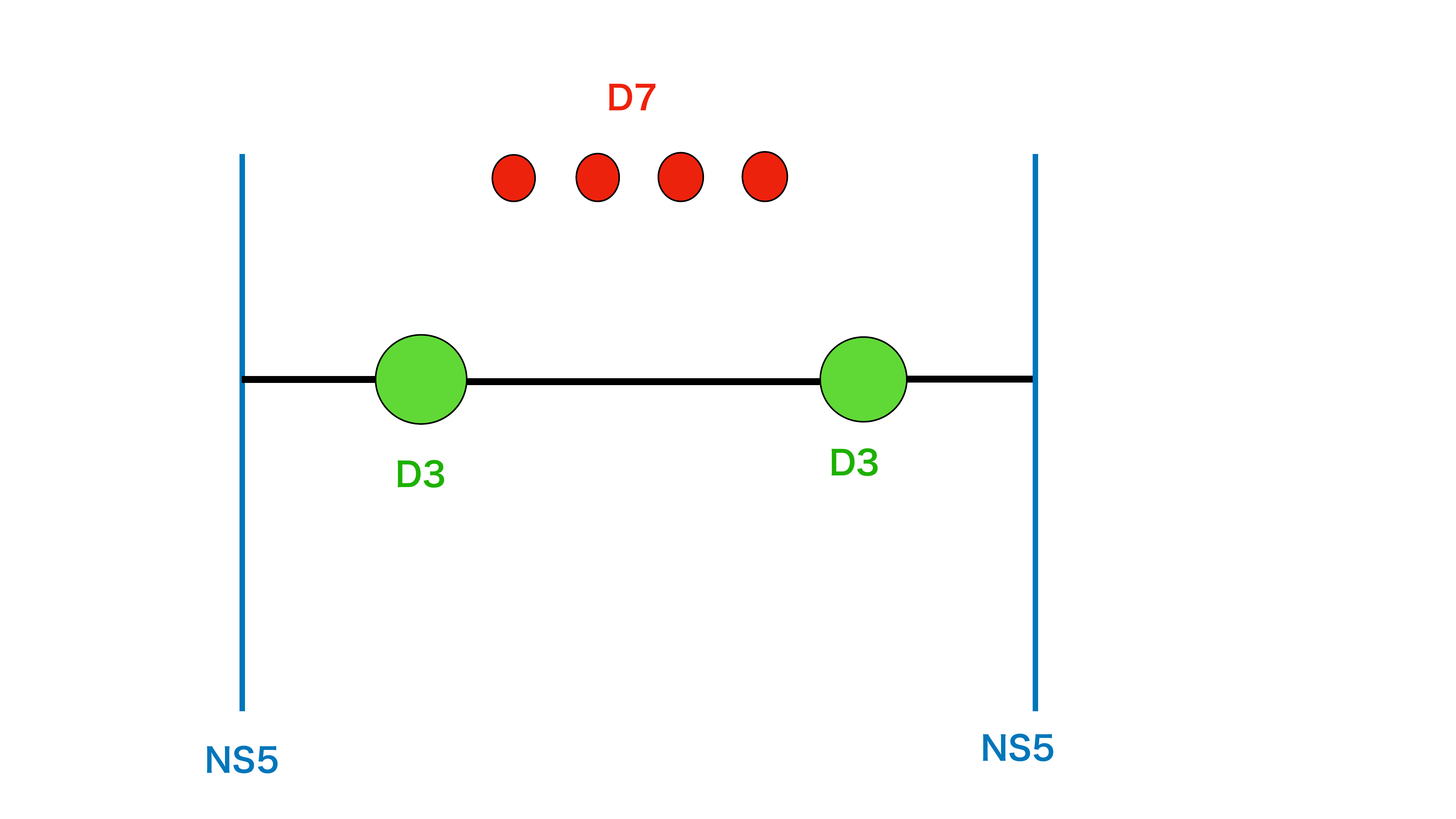}
}
\subfigure[]{\label{fig:subbrane4}
\includegraphics[height=4cm]{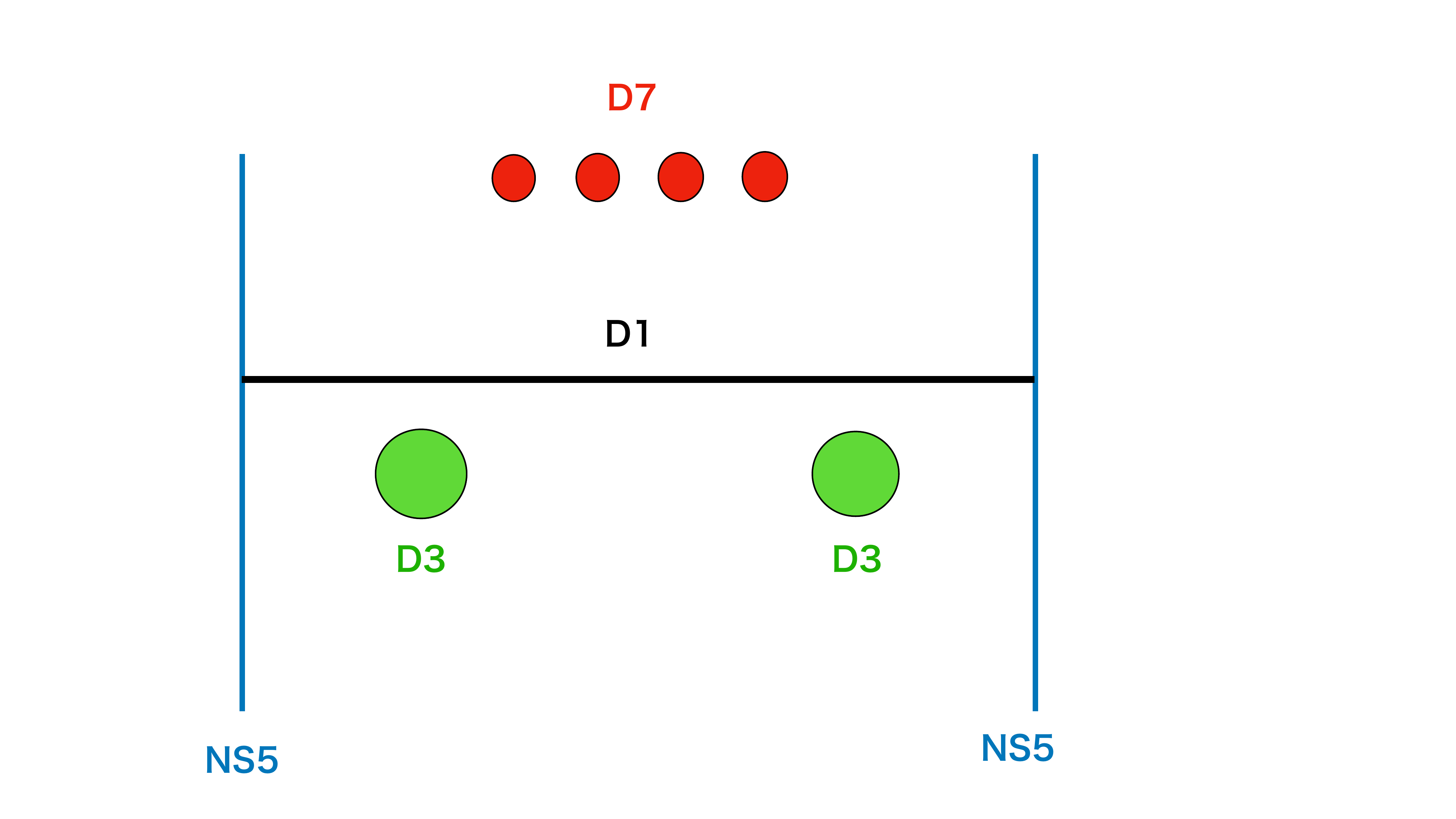}}
\caption{(a): A D1-brane suspended between two D3-branes is added to Figure \ref{fig:subbrane2}. 
(b): Three segments of D1-branes form a single D1-brane, which ends on the NS5-branes but not on the D3-branes. This screens the D1-brane charge, providing a D-brane realization of monopole bubbling.}
\label{fig:brane2}
\end{figure}
\end{pdffig}

We briefly review the naive brane picture for monopole bubbling of 't Hooft loops in 4d \(\mathcal{N}=2\) \(U(N)\) and \(SU(N)\) gauge theories with \(2N\) hypermultiplets, as proposed in \cite{Brennan:2018rcn}; see also \cite{Brennan:2018yuj}, as well as its completion through the introduction of extra D5-branes \cite{Assel:2019iae}. Our focus is on the case \(N=2\), i.e., the \(SU(2)\) gauge theory with four hypermultiplets. The advantage of the complete D-brane setup with extra D5-branes is that it avoids complicated calculations in the Born--Oppenheimer approximation and can also be applied to dyonic loops. However, so far, the complete D-brane setup is known only for the gauge group \(SU(N)\).

The brane configuration is depicted in Table \ref{table:brane}. In type IIB string theory, we introduce \(N\) D3-branes extending along the \(x^0, x^1, x^2, x^3\) directions. The low-energy worldvolume theory on the D3-branes is a 4d \(\mathcal{N}=4\) \(U(N)\) supersymmetric gauge theory. 
As in \cite{Assel:2019iae}, the D3-branes are placed at the origin of the $x^{6,7,8,9}$-
space, which is the fixed point of an $\Omega$-background   acting on the $x^{6,7,8,9}$-directions.  This $\Omega$-background  gives a mass to the adjoint hypermultiplet and realizes the corresponding $\mathcal{N}=2^*$ deformation; the relation between $\mathcal{N}=2^*$ mass and this $\Omega$-parameter is described below.
We also introduce \(2N\) D7-branes extending along the \(x^i\) directions for \(i=0,1,2,3,6,7,8,9\), which provide \(2N\) fundamental hypermultiplets. Sending the adjoint mass to infinity decouples the adjoint hypermultiplet and leaves the $\mathcal{N}=2$ $U(N)$ gauge theory with $2N$
 fundamental hypermultiplets.

An 't Hooft loop is realized by NS5-branes extending along the \(x^i\) directions for \(i=0, 5, 6, 7, 8, 9\). 
An NS5-brane placed between the \(i\)-th D3-brane and the \((i+1)\)-th D3-brane (counting from the left) gives rise to a 't Hooft loop with magnetic charge 
${\bm h}_i = {\rm diag} (\frac{1}{2}, \cdots, \underset{i}{\frac{1}{2}}, \underset{i+1}{-\frac{1}{2}}, \cdots, -\frac{1}{2} ) \in \mathfrak{u}(N)$.
An 't Hooft loop with magnetic charge \({\bm p}=\sum_{i=1}^N n_i {\bm h}_i\) is obtained by placing \(n_i\) NS5-branes between the \(i\)-th and \((i+1)\)-th D3-branes for \(i=1, \dots, N\). The 't Hooft loops in the \(SU(N)\) gauge theory are obtained by restricting the magnetic charge \({\bm p}\) to belong to the coweight lattice of \(\mathfrak{su}(N)\).

For example, Figure \ref{fig:subbrane1} depicts a brane configuration for an 't Hooft loop \(L_{({\bm p}, {\bm 0})}\) with \({\bm p}=(1,-1)\) in \(U(2)\) or \(SU(2)\) gauge theory with four hypermultiplets.\footnote{A magnetic charge \({\bm p}=(p,-p)\) with \(p \in \mathbb{Z}\) in the \(U(2)\) gauge theory corresponds to \({\bm p}=p\) in the \(SU(2)\) gauge theory.} Figure \ref{fig:subbrane2} depicts another configuration for the 't Hooft loop \(L_{({\bm p}, {\bm 0})}\) with \({\bm p}=(1,-1)\), related to Figure \ref{fig:subbrane1} by the Hanany--Witten effect.

The physical interpretation of monopole bubbling is that an 't Hooft--Polyakov monopole near the 't Hooft loop screens the magnetic charge \({\bm p}\) of the 't Hooft loop. The brane realization of this phenomenon is as follows.
An 't Hooft--Polyakov monopole with a magnetic charge \((0,\dots,0,\underset{i}{-1},\underset{i+1}{1},0,\dots,0)\) is realized by a D1-brane stretched between the \(i\)-th and \((i+1)\)-th D3-branes \cite{Diaconescu:1996rk}; see Figure \ref{fig:subbrane3} for the \(N=2\) case.
When D1-branes corresponding to an 't Hooft loop and a D1-brane corresponding to an 't Hooft--Polyakov monopole form a single D1-brane, the resulting D1-brane no longer ends on the D3-branes, but instead ends on the two NS5-branes, as depicted in Figure \ref{fig:subbrane4}.
In this case, the magnetic charge of the 't Hooft loop is screened. The proposal in \cite{Brennan:2018rcn, Brennan:2018yuj} is that \(Z_{\rm mono}\) is given by the Witten index of the worldvolume theory on the D1-branes suspended between the NS5-branes.

\begin{pdffig}
\begin{figure}[thb]
\centering
\subfigure[]{\label{fig:subquiver1}
\includegraphics[height=3.7cm]{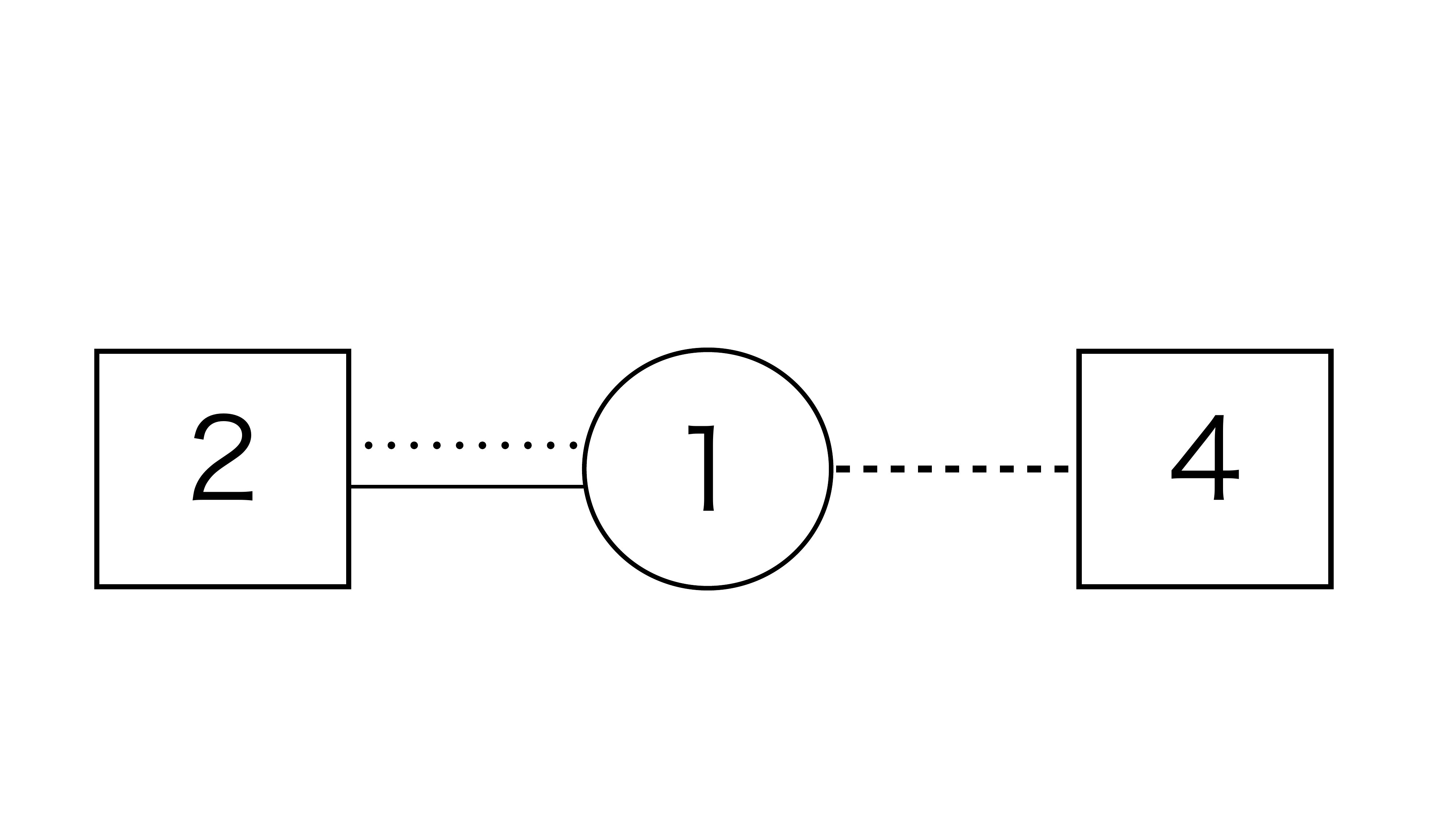}
}
\subfigure[]{\label{fig:subquiver2}
\includegraphics[height=3.7cm]{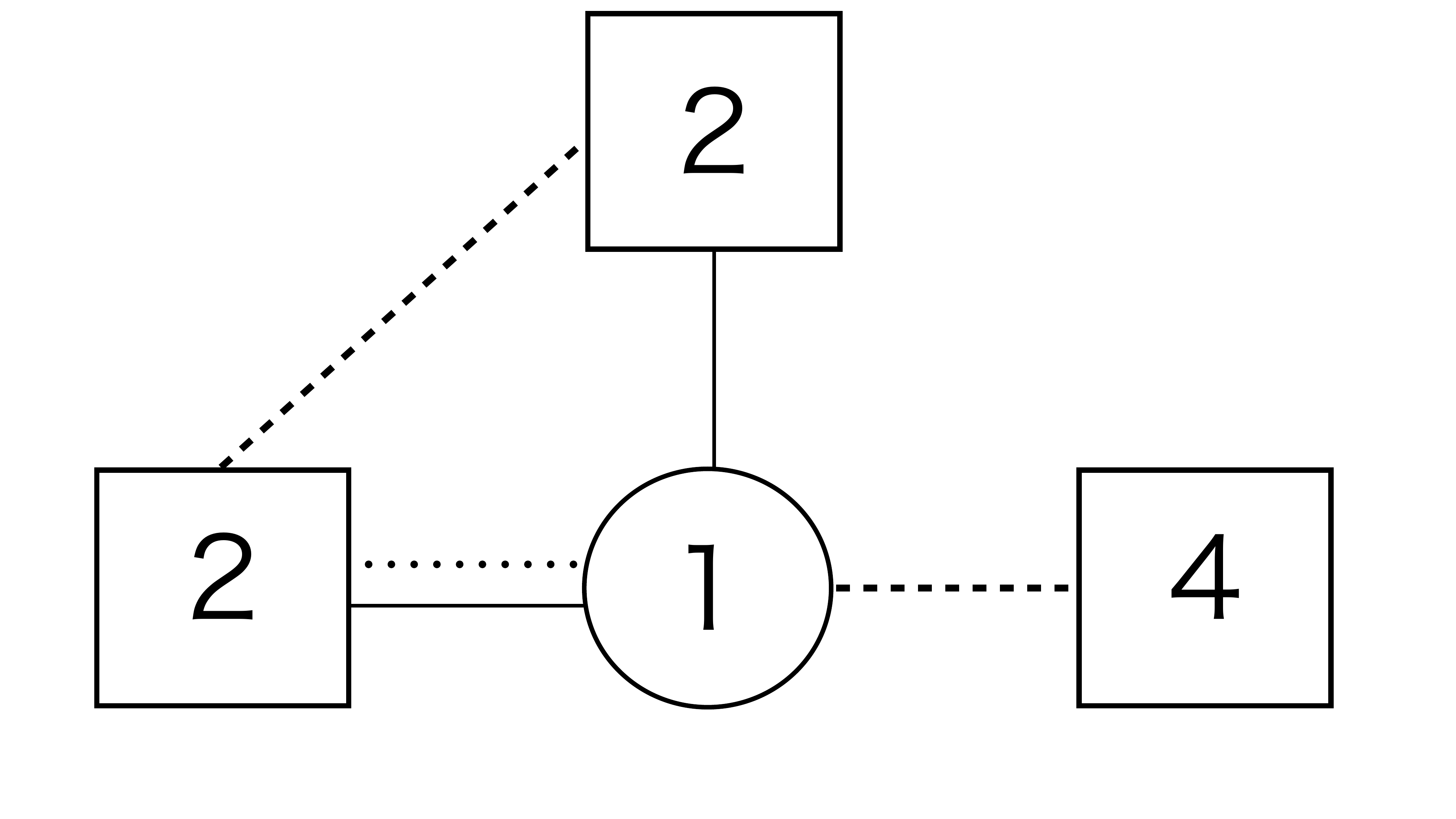}}
\caption{(a): The quiver diagram representing the 1d supermultiplets associated with the D1-brane worldvolume theory in Figure \ref{fig:subbrane4}.
(b): The quiver diagram representing the 1d supermultiplets associated with the D1-brane worldvolume theory in Figure \ref{fig:subbrane8}. 
The circle represents the \(U(1)\) vector multiplet. The solid and dotted lines represent \(\mathcal{N}=(4,4)\) hypermultiplets. The dashed line represents \(\mathcal{N}=(0,4)\) Fermi multiplets. The number in a box indicates the number of supermultiplets represented by the line attached to the box.}
\label{fig:quiver1}
\end{figure}
\end{pdffig}

Let us evaluate \(Z^{(\zeta)}_{\rm JK}\) in the \(SU(2)\) gauge theory associated with the brane configuration depicted in Figure \ref{fig:subbrane4}. 
The quiver diagram representing the matter content of the SQM is shown in Figure \ref{fig:subquiver1}. 
The 1d \(U(1)\) vector multiplet, \(\mathcal{N}=(4,4)\) hypermultiplets, and \(\mathcal{N}=(0,4)\) Fermi multiplets arise from D1-D1 strings, D1-D3 strings, and D1-D7 strings, respectively.\footnote{In this article, we allow a slight abuse of notation and let \(\mathcal{N}=(s,t)\) denote the dimensional reduction of 2d \(\mathcal{N}=(s,t)\) supersymmetry to one dimension.}

The SQM has the global symmetry $SU(2)_+ \times SU(2)_1 \times SU(2)_2$, with  
chemical potentials $-2 \epsilon_+, \epsilon_1,$ and $\epsilon_2$, respectively. 
In the brane picture these factors are identified with ten-dimensional spacetime rotations $SO(3)_{123} \simeq SU(2)_{+}$ and $SO(4)_{6789} \simeq SU(2)_1 \times SU(2)_2$.
The chemical potentials $-2 \epsilon_+, \epsilon_1,$ and $\epsilon_2$ are the $\Omega$-background (equivariant) parameters for rotations in the $x^{2,3}$, $x^{6,7}$, and $x^{8,9}$ planes, respectively \cite{Assel:2019iae}. Supersymmetry requires the combined rotations in the \(x^{2,3}\), \(x^{6,7}\), and \(x^{8,9}\) planes to leave the preserved supercharge invariant. This imposes
\begin{align}
\epsilon_+=\frac{\epsilon_1+\epsilon_2}{2}\,.
\end{align}
On the other hand, we define 
\begin{align}
m_*:=\frac{\epsilon_1-\epsilon_2}{2}\,.
\end{align}
The corresponding  rotation   in the $x^{6,7,8,9}$-directions acts as $(x^6 +{\rm i} x^{7}, x^8 +{\rm i} x^{9}) \to ( e^{{\rm i} \alpha }(x^6 +{\rm i} x^{7}),e^{-{\rm i}\alpha }( x^8 +{\rm i} x^{9})) $. On the D3-brane worldvolume, this rotation is identified with 
 the $U(1)$ flavor symmetry  acting on the adjoint hypermultiplet, and $m_*$ is its chemical potential. Therefore,  $m_*$ is identified with the $\mathcal{N}=2^*$ mass of the adjoint hypermultiplet. The limit $m_* \to \infty$ therefore decouples the adjoint hypermultiplet.

The Witten index with a nonzero FI parameter \(\zeta\) is evaluated using the SUSY localization formula \cite{Hori:2014tda, Hwang:2014uwa} as
\begin{align}
&Z^{(\zeta)}_{\rm JK}({\bm p}=1, \tilde{\bm p}=0, {\bm q}=0 ) \nonumber \\
&= \lim_{m_* \to \infty}
\oint_{{\rm JK}(\zeta)} \frac{du}{2 \pi {\rm i}} \frac{-{\rm sh}(2\epsilon_+)}{{\rm sh}(m_*)^2 {\rm sh}(\epsilon_1){\rm sh}(\epsilon_2)}
\times \prod_{i=1}^2 \frac{{\rm sh}(\pm(u-a_i) +m_*)}{{\rm sh}(\pm(u-a_i) +\epsilon_+)}
 \prod_{f=1}^{4} {\rm sh}(u-m_f) 
  \\
&=\oint_{{\rm JK}(\zeta)} \frac{d u}{2 \pi {\rm i} }  \frac{ \sh (2\epsilon_+) \prod_{f=1}^4  \sh (  u - m_f)}{\prod_{i=1}^2 \sh ( \pm(u - a_i)+\epsilon_+)  }\,.
\label{eq:monopoleSU2}
\end{align}
Here, \( f(\pm x):=\prod_{s=\pm1}f(s x) \), and \((a_1, a_2)=(a,-a)\).
The limit \(\lim_{m_* \to \infty}\) corresponds to taking the \(\mathcal{N}=2^*\) mass to infinity. 
The factor \({\rm sh}(m_*)^2\) in the denominator is introduced for regularization.
\(\oint_{{\rm JK}(\zeta)}\) denotes the Jeffrey--Kirwan (JK) residue.  
In this case, the JK residue is taken at the following poles:
\begin{align}
 u= \left\{
\begin{aligned}
  \pm a-\epsilon_+\, &\text{ for } \zeta>0, \\
 \pm a+\epsilon_+\, &\text{ for } \zeta<0.
\end{aligned}
\right.
\end{align}
Then, \(Z^{(\zeta)}_{\rm JK}\) is given by
\begin{align}
 Z^{(\zeta)}_{\rm JK} = \left\{
\begin{aligned}
\frac{\prod_{f=1}^4 \sh(a-m_f-\epsilon_+)}{\sh(2a) \sh(-2a+2\epsilon_+) }+\frac{\prod_{f=1}^4 \sh(-a-m_f-\epsilon_+)}{\sh(-2a) \sh(2a+2\epsilon_+) } & \,\text{ for } \zeta > 0, \\
\frac{\prod_{f=1}^4 \sh(a-m_f+\epsilon_+)}{\sh(-2a) \sh(2a+2\epsilon_+) }+\frac{\prod_{f=1}^4 \sh(-a-m_f+\epsilon_+)}{\sh(2a) \sh(-2a+2\epsilon_+) } & \,\text{ for } \zeta < 0.
\end{aligned}
\right.
\label{eq:JKp}
\end{align}
Note that the Witten index \eqref{eq:JKp} evaluated in the two chambers exhibits a wall-crossing phenomenon: \(Z^{(\zeta>0)}_{\rm JK} \neq Z^{(\zeta<0)}_{\rm JK}\).

\subsection{Complete brane setup with extra D5-branes}
\label{sec:imp}

\begin{pdffig}
\begin{figure}[thb]
\centering
\subfigure[]{\label{fig:subbrane7}
\includegraphics[height=4cm]{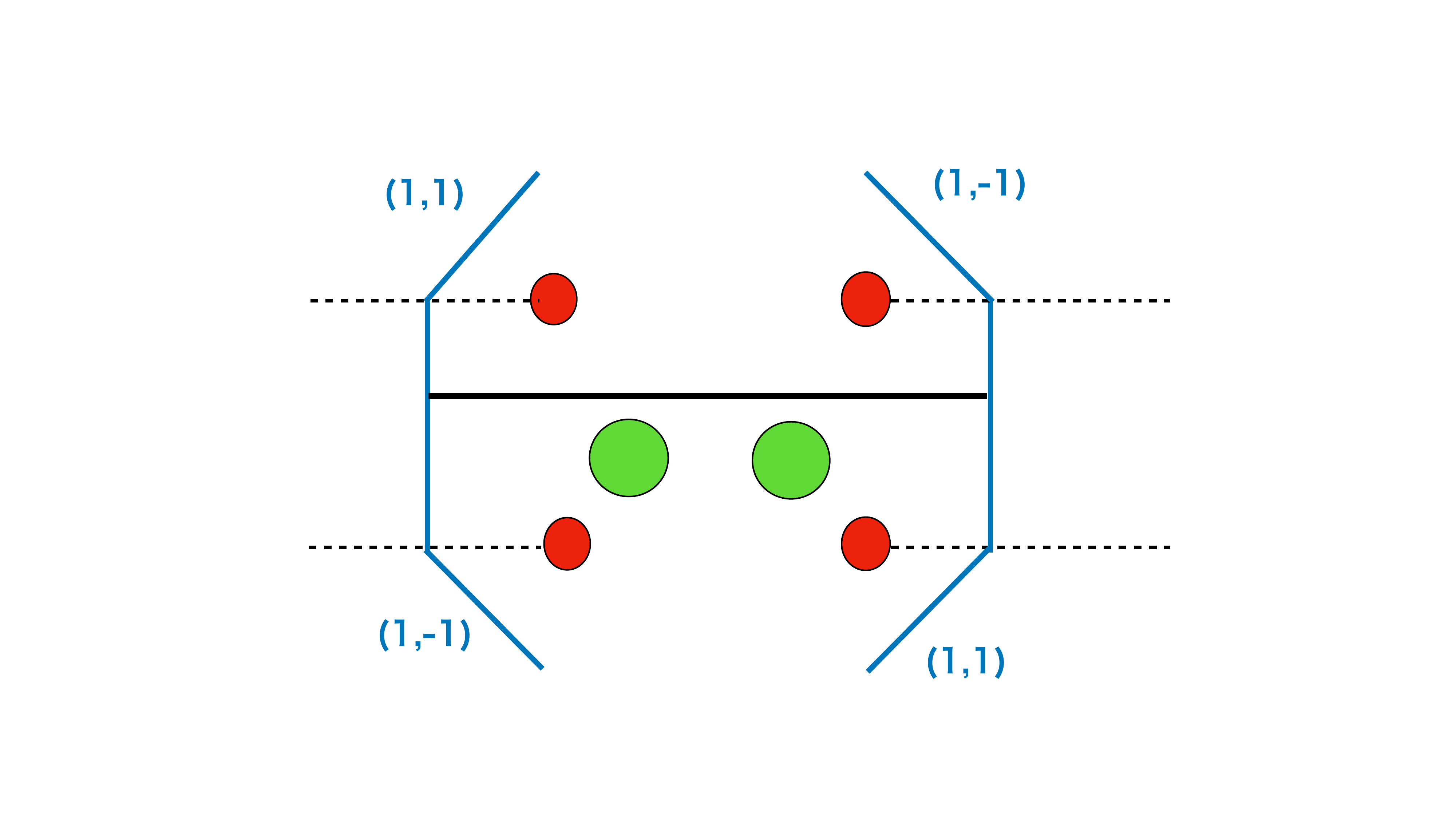}
}
\subfigure[]{\label{fig:subbrane8}
\includegraphics[height=4cm]{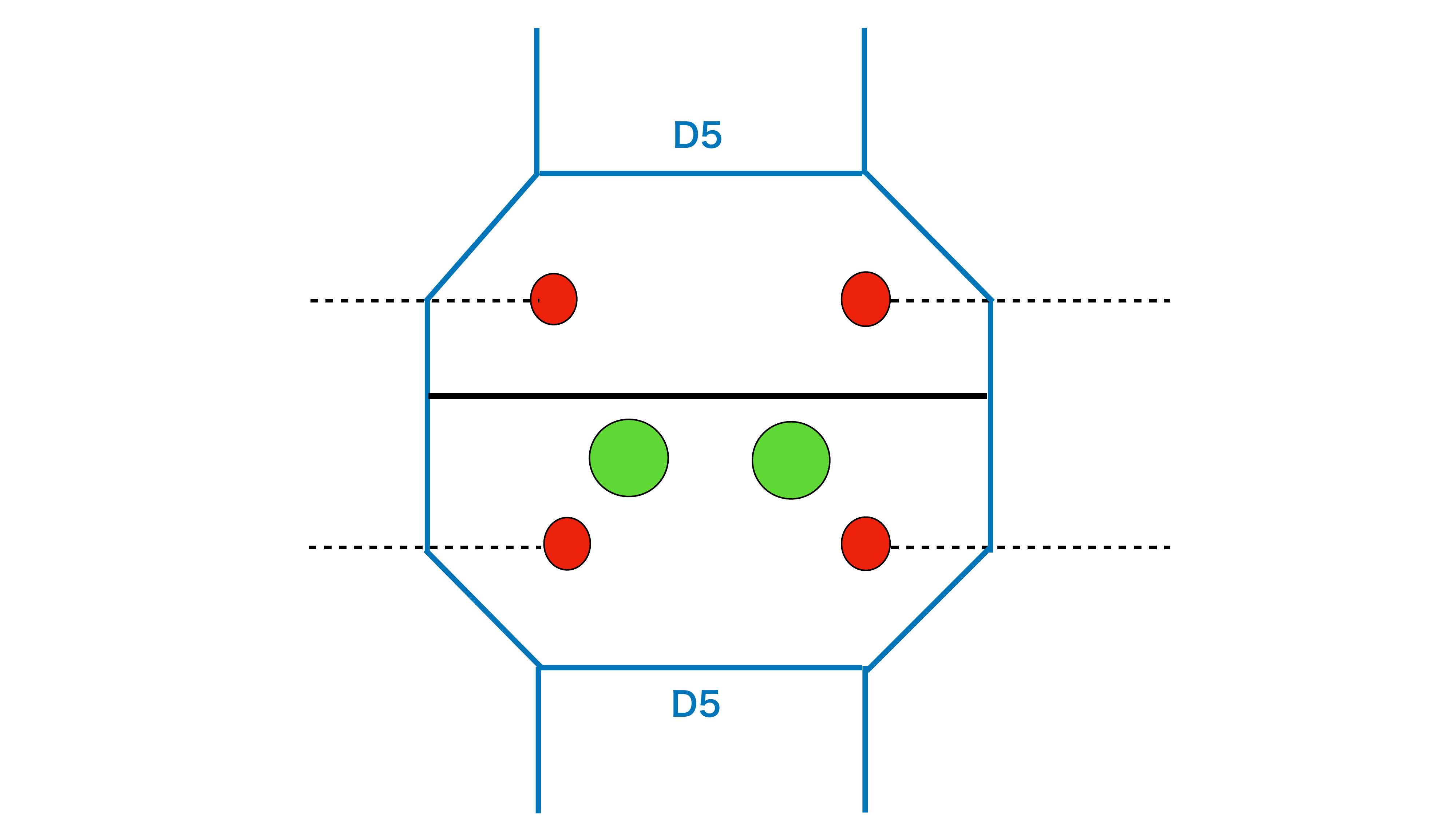}}
\caption{(a): When NS5-branes cross the branch cuts (denoted by black dashed lines) associated with D7-branes, \((1,\pm1)\) 5-branes (depicted as four oblique blue lines) are created. 
Note that the semi-infinite \((1,1)\) and \((1,-1)\) 5-branes intersect at two points, one above and one below, in the \((x^4, x^5)\) plane. 
(b): The improved brane configuration for monopole bubbling.
Two D5-branes (denoted by two horizontal blue lines) are introduced. Due to charge conservation at the junctions, the semi-infinite 5-branes are converted into NS5-branes.}
\label{fig:brane4}
\end{figure}
\end{pdffig}

In the previous subsection, the bending effect of 5-branes in the presence of D7-branes was not taken into account.
For example, in Figure \ref{fig:subbrane7}, semi-infinite \((1, \pm1)\) 5-branes appear due to the branch cuts associated with D7-branes.
Since these semi-infinite 5-branes intersect in the \((x^4, x^5)\) plane, the matter content derived from the naive brane configuration in the previous subsection does not capture this effect \cite{Assel:2019iae}.
The authors of \cite{Assel:2019iae} proposed completing the brane configuration by adding extra D5-branes.
Then, the monopole bubbling contribution \(Z_{\rm mono}\) is obtained by considering the zero-flavor-charge sector in the Witten index associated with the extra D5-branes, where the zero-flavor-charge sector corresponds to taking the extra D5-branes to infinity in the \((x^4,x^5)\) plane.

An important point here is that the modified SQM with extra D5-branes does not exhibit wall crossing. The Witten index of the modified SQM can be reliably calculated using the JK-residue method with a particular choice of the FI parameter.

The complete brane setup for \({\bm p}=(1,-1)\) (resp. \({\bm p}=1\)) and \(\tilde{\bm p}=(0,0)\) (resp. \(\tilde{\bm p}=0\)) in the \(U(2)\) (resp. \(SU(2)\)) gauge theory is depicted in Figure \ref{fig:subbrane8}. 
A new neutral Fermi multiplet arising from D5-D3 strings is added to the SQM depicted in Figure \ref{fig:subquiver1}.
Thus, the modified SQM with extra D5-branes is depicted in Figure \ref{fig:subquiver2}.
The two extra D5-branes are taken to infinity in the final step of the calculation.

The monopole bubbling contribution is evaluated as
\begin{align}
Z_{\rm mono} &=
\lim_{w_2 \to \infty}  \lim_{w_1 \to 0} \lim_{m_* \to \infty}
\oint_{{\rm JK}(\zeta)} \frac{du}{2 \pi {\rm i}} \frac{(-1){\rm sh}(2\epsilon_+)}{{\rm sh}(m_*)^2{\rm sh}(\epsilon_1){\rm sh}(\epsilon_2)}
 \prod_{i=1}^2 \frac{{\rm sh}(\pm(u-a_i) +m_*)}{{\rm sh}(\pm(u-a_i) +\epsilon_+)}
  \nonumber \\
& \qquad \times  \prod_{k=1}^{4} {\rm sh}(u-m_k) \cdot
 \prod_{n=1}^{2} \frac{\prod_{i=1}^2 {\rm sh}(a_i-v_n)}{ {\rm sh}(\pm(u-v_n) -\epsilon_+) } 
\label{eq:monohooft} \\
&=\frac{\prod_{f=1}^4 \sh(a-m_f-\epsilon_+)}{\sh(2a) \sh(-2a+2\epsilon_+) }+\frac{\prod_{f=1}^4 \sh(-a-m_f-\epsilon_+)}{\sh(-2a) \sh(2a+2\epsilon_+) }
+\ch\Bigl(\sum_{f=1}^4 m_f +2 \epsilon_+\Bigr)\,.
\label{eq:monothooft}
\end{align}
Here, \(\ch(x):=2\cosh(x/2)\), \(w_{l}:=e^{-v_l}\) for \(l=1,2\), and \(v_l\) are the flavor fugacities associated with the hypermultiplets and Fermi multiplets corresponding to the extra D5-branes.
The limit \(\lim_{w_2 \to \infty}  \lim_{w_1 \to 0}\) 
corresponds to taking one D5-brane to positive infinity and the other to negative infinity in the \((x^4, x^5)\) plane.

Note that the first and second terms in \eqref{eq:monothooft} are exactly the same as \(Z^{(\zeta >0)}_{\rm JK}\) in \eqref{eq:JKp}, while the third term is the extra term \(Z^{(\zeta>0)}_{\rm extra}\).
Here, we evaluate the JK residue in the positive FI parameter region: \(\zeta>0\). In the negative FI parameter region, \(Z_{\rm mono}\) is obtained by the sign flip
of the \(\Omega\)-background parameter: \(\epsilon_+ \mapsto - \epsilon_+\) in \eqref{eq:monothooft}.
The sign flip is expressed as a combination of a reflection in the \(x^3\) direction and an R-symmetry transformation. Since elementary BPS loop operators are invariant under reflection, this symmetry provides an important consistency check of the localization computation \cite{Assel:2019iae}. Although the expressions for \(Z_{\rm mono}\) in the positive and negative FI parameter regions appear different, they are actually equivalent.
On the other hand, when an 't Hooft loop is decomposed into a product of 't Hooft loops with smaller magnetic charges, the sign flip corresponds to a change in the ordering of the operator product \cite{Tachikawa:2015iba}. In this case, the ordering of 't Hooft loops correlates with the wall-crossing phenomenon of \(Z^{(\zeta)}_{\rm JK}\) \cite{Hayashi:2019rpw}.

\subsection{Monopole bubbling effect for dyonic loops}
\label{sec:dyonbub}

\begin{pdffig}
\begin{figure}[thb]
\centering
\subfigure[]{\label{fig:subbrane9}
\includegraphics[height=4.1cm]{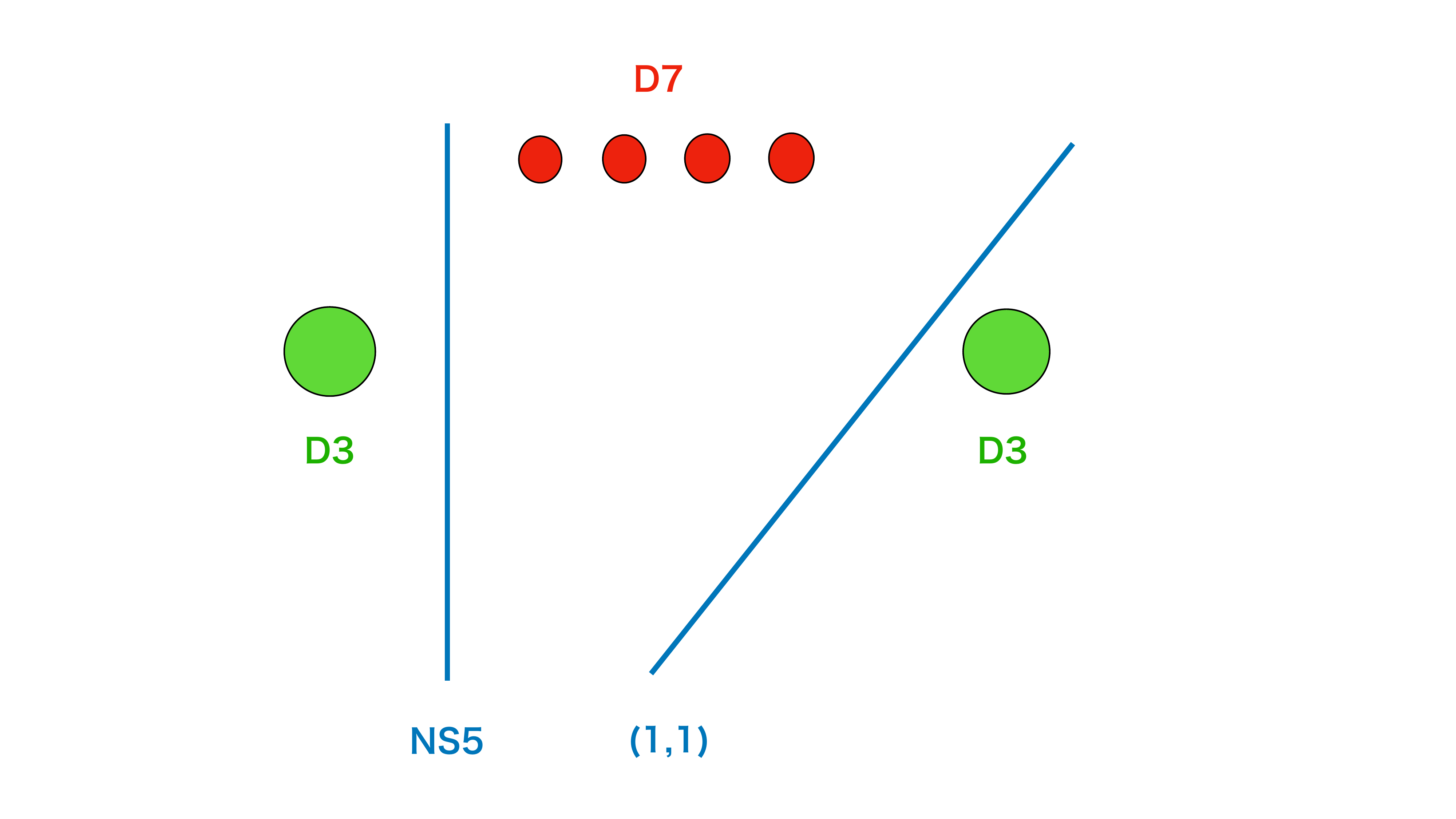}
}
\subfigure[]{\label{fig:subbrane10}
\includegraphics[height=4.8cm]{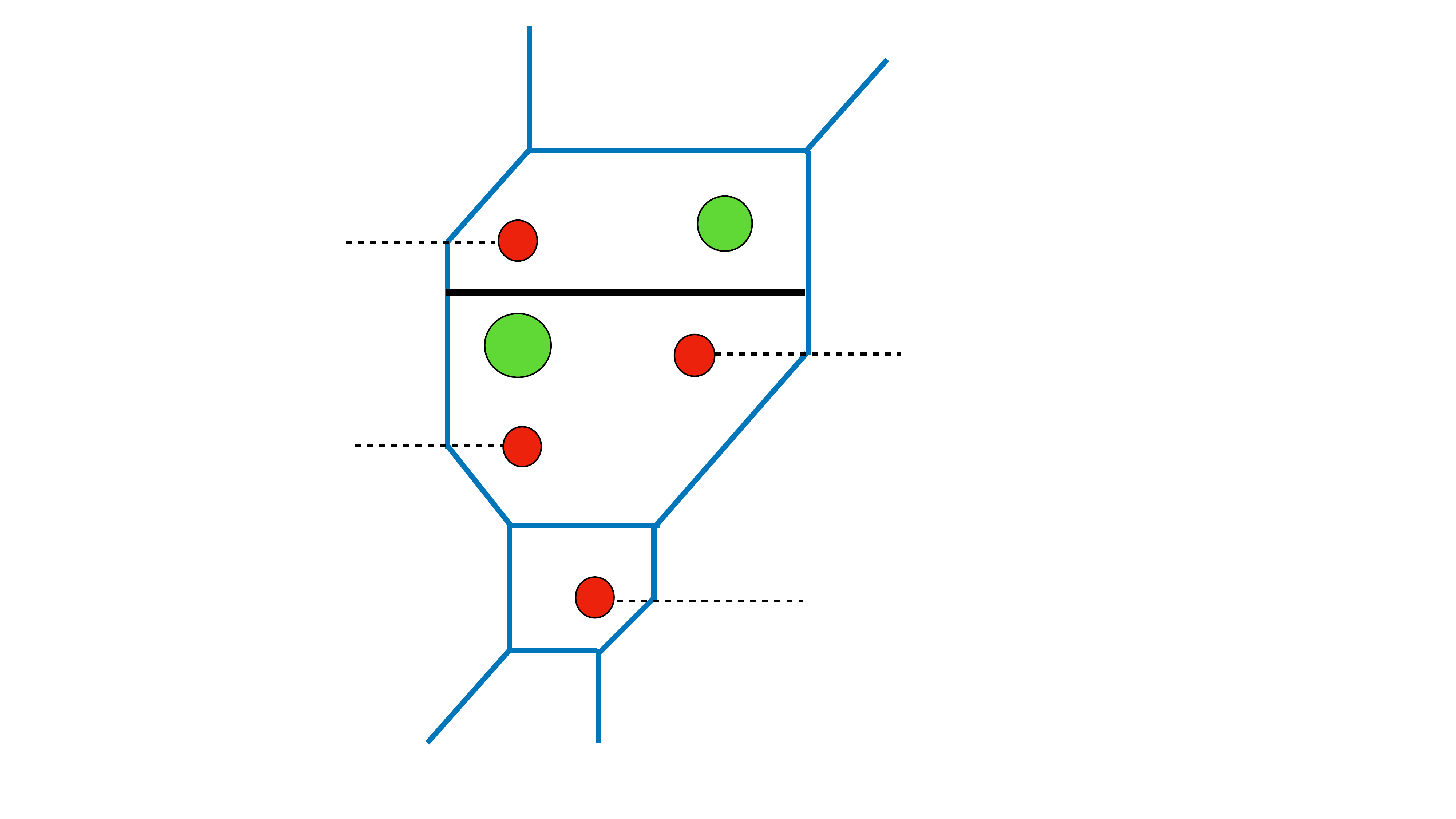}}
\caption{(a): The brane configuration for a dyonic loop with \({\bm p}=1\) and \({\bm q}=1\). 
(b): The brane configuration for monopole bubbling in the dyonic loop with \(\tilde{\bm p}=0\).}
\label{fig:brane41}
\end{figure}
\end{pdffig}

\begin{pdffig}
\begin{figure}[t]
\centering
\includegraphics[height=4cm]{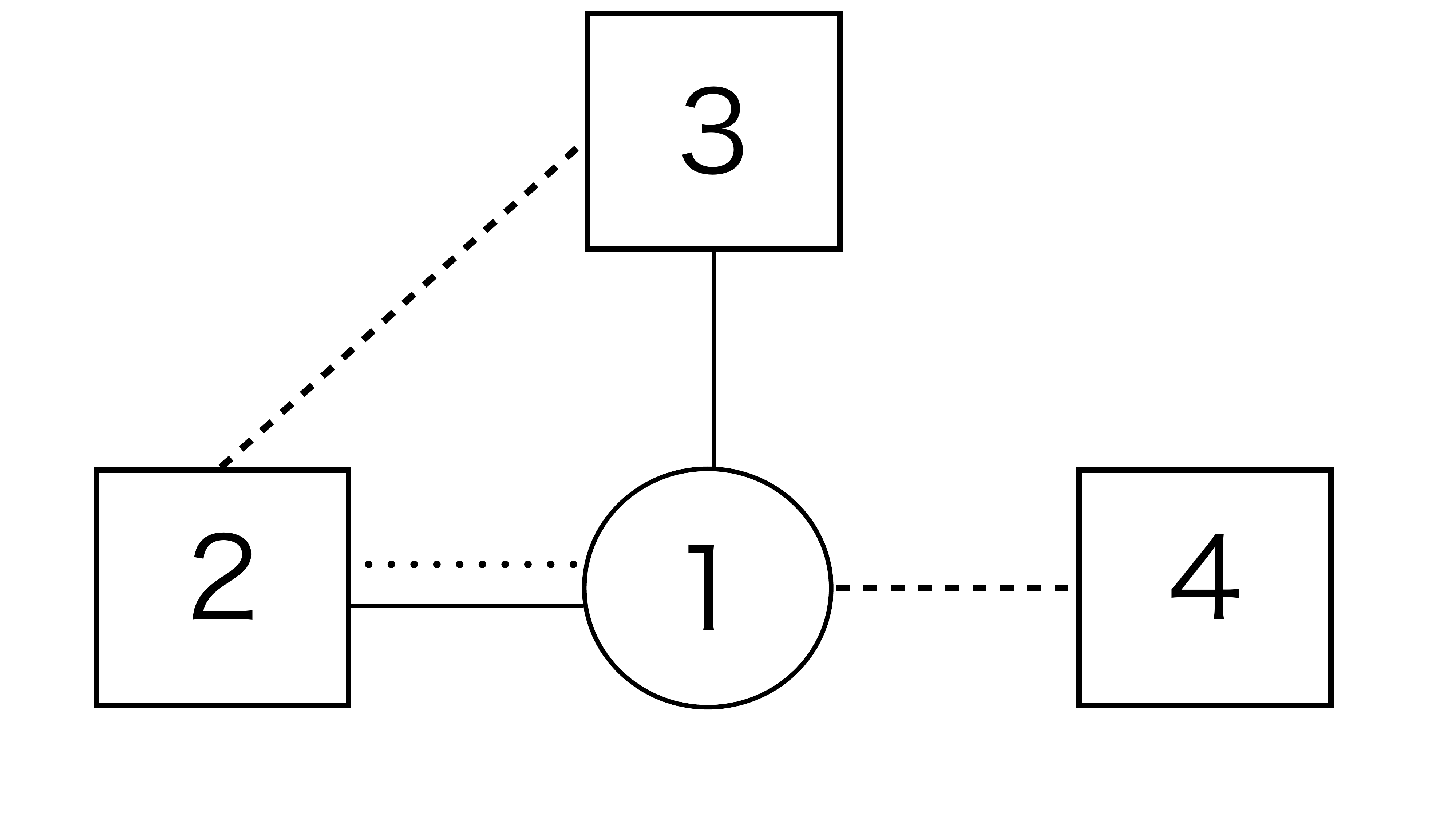}
\caption{The quiver diagram of the SQM for monopole bubbling in the dyonic loop \(L_{(1,1)}\).}
\label{fig:quiver3}
\end{figure}
\end{pdffig}

Next, we consider the monopole bubbling effect for a dyonic loop \(L_{(1,1)}\) in the \(SU(2)\) gauge theory, following the brane setup in \cite{Assel:2019iae}. The D-brane configuration for \(L_{(1,1)}\) is obtained by replacing an NS5-brane with a \((1,1)\) 5-brane in the setup for the 't Hooft loop \(L_{(1,0)}\); see Figure \ref{fig:subbrane9}.  
The complete brane configuration for monopole bubbling is achieved by introducing a D1-brane between two D3-branes, along with three extra D5-branes. After the Hanany--Witten effect occurs, we obtain the brane configuration depicted in Figure \ref{fig:subbrane10}. The quiver diagram of the SQM is shown in Figure \ref{fig:quiver3}. As before, the monopole bubbling contribution is obtained from the neutral charge sector associated with the extra D5-branes in the Witten index:
\begin{align}
&Z_{\rm mono}({\bm p}=1, \tilde{\bm p}=0, {\bm q}=1) \nonumber \\
&= 
\lim_{w_3 \to \infty}  \lim_{w_1,w_2 \to 0} \lim_{m_* \to \infty} \oint_{{\rm JK}(\zeta)} \frac{du}{2 \pi {\rm i}} \frac{(-1){\rm sh}(2\epsilon_+)}{{\rm sh}(m_*)^2 {\rm sh}(\epsilon_1){\rm sh}(\epsilon_2)}
 \prod_{i=1}^2 \frac{{\rm sh}(\pm(u-a_i) +m_*)}{{\rm sh}(\pm(u-a_i) +\epsilon_+)}
 \nonumber \\
& \qquad \times \prod_{k=1}^{4} {\rm sh}(u-m_k) 
  \prod_{n=1}^{3} \frac{\prod_{i=1}^2 {\rm sh}(a_i-v_n)}{ {\rm sh}(\pm(u-v_n) -\epsilon_+) } 
 \\
&= e^{-a +\epsilon_+}\frac{\prod_{f=1}^4 \sh(a-m_f-\epsilon_+)}{\sh(2a) \sh(-2a+2\epsilon_+) }
+ e^{a +\epsilon_+} \frac{\prod_{f=1}^4 \sh(-a-m_f-\epsilon_+)}{\sh(-2a) \sh(2a+2\epsilon_+) }
\nonumber \\
&\qquad 
-e^{\frac{1}{2}(2\epsilon_+ +\sum_{f=1}^4 m_f)} \Bigl(\sum_{f=1}^4 e^{-m_f} - e^{\epsilon_+-a}- e^{\epsilon_+ +a}\Bigr)\,.
\label{eq:dyonbub}
\end{align}
As we will see,  the first line of \eqref{eq:dyonbub} agrees with the VEV of a 1d BPS Wilson loop,
\[
\langle e^{{\rm i}  \oint (A^{(1d)}_t -{\rm i} \sigma^{(1d)}) dt } \rangle^{(\zeta)}_{\rm SQM},
\]
in the SQM with \(\zeta \neq 0\), described by the quiver diagram in Figure \ref{fig:subquiver1}. Here, \(A^{(1d)}_t\) and \(\sigma^{(1d)}\) are the gauge and scalar fields in the 1d \(\mathcal{N}=(0,2)\) \(U(1)\) vector multiplet. The VEV of this Wilson loop is computed via the localization formula:
\begin{align}
\langle e^{{\rm i}  \oint (A^{(1d)}_t -{\rm i} \sigma^{(1d)}) dt } \rangle^{(\zeta)}_{\rm SQM}
&=\oint_{{\rm JK}(\zeta)} \frac{d u}{2 \pi {\rm i} } e^{u} \frac{ \sh (2\epsilon_+) \prod_{f=1}^4  \sh (  u - m_f)}{\prod_{i=1}^2 \sh ( \pm(u - a_i)+\epsilon_+)  }\,.
\label{eq:naiveWilson}
\end{align}
Here, the factor \(e^{u}\) corresponds to the insertion of the Wilson loop in \eqref{eq:monopoleSU2}. When \(\zeta>0\), the JK residue is evaluated as
\begin{align}
\langle e^{{\rm i}  \oint (A^{(1d)}_t -i {\rm }\sigma^{(1d)}) dt } \rangle^{(\zeta>0)}_{\rm SQM}
 &=
e^{-a +\epsilon_+}\frac{\prod_{f=1}^4 \sh(a-m_f-\epsilon_+)}{\sh(2a) \sh(-2a+2\epsilon_+) }
+ e^{a +\epsilon_+} \frac{\prod_{f=1}^4 \sh(-a-m_f-\epsilon_+)}{\sh(-2a) \sh(2a+2\epsilon_+) }\,.
\label{eq:JKWilson}
\end{align}
Thus, \eqref{eq:JKWilson} exactly matches the first line of \eqref{eq:dyonbub}. The JK residue in the negative FI parameter region \(\zeta <0\) is obtained by flipping the sign of the \(\Omega\)-background parameter: \(\epsilon_+ \mapsto -\epsilon_+\) in \eqref{eq:JKWilson}.

Although the 't Hooft loop has a brane setup before completion, the dyonic loop does not. As a result, \(Z^{(\zeta)}_{\rm JK}\) for the dyonic loop lacks a clear brane interpretation. Since the VEV of the 1d Wilson loop is computed by the JK residue of the SQM, and since \eqref{eq:JKWilson} exactly matches the first line of \eqref{eq:dyonbub}, it is natural to interpret the VEV of the 1d Wilson loop with $\zeta \neq 0$ as the JK part of the monopole bubbling effect for the dyonic loop, namely \(Z^{(\zeta)}_{\rm JK}({\bm p}=1, \tilde{\bm p}=0, {\bm q}=1)\). Then, the second line of \eqref{eq:dyonbub} is identified with the extra term \(Z^{(\zeta)}_{\rm extra}({\bm p}=1, \tilde{\bm p}=0, {\bm q}=1)\).

\subsection{$Z^{(\zeta)}_{\rm extra}$ as the decoupled states in $Z^{(\zeta)}_{\rm JK}$}
\label{sec:subt}

In \cite{Hayashi:2020ofu}, it was pointed out that the extra term \(-Z^{(\zeta)}_{\rm extra}\) in the 't Hooft loop with minimal magnetic charge is given by the states in \(Z^{(\zeta)}_{\rm JK}\) that are neutral under the 4d global gauge symmetry, or equivalently, independent of \(e^{-a}\). These decoupled states, which are irrelevant to the four-dimensional dynamics, must be removed from \(Z^{(\zeta)}_{\rm JK}\)\footnote{A similar phenomenon is also known in 5d gauge theories, where the correct answer is obtained by separating decoupled states from Nekrasov's instanton partition function \cite{Hayashi:2013qwa, Bao:2013pwa}.}.
For example, \(Z^{(\zeta>0)}_{\rm JK}({\bm p}=1, \tilde{\bm p}=0, {\bm q}=0)\) can be expanded as
\begin{align}
Z^{(\zeta>0)}_{\rm JK}({\bm p}=1, \tilde{\bm p}=0, {\bm q}=0)&=-\ch\Bigl(\sum_{f=1}^4 m_f +2 \epsilon_+\Bigr)+\sum_{n=1}^{\infty} c_n ({\bm m}, \epsilon_+)  e^{-n a}.
\label{eq:expand1}
\end{align}
The first term in \eqref{eq:expand1} is independent of \(a\) and reproduces \(-Z^{(\zeta)}_{\rm extra}\).

Next, we observe that this expansion method can also be applied to the dyonic loop. Expanding \(Z^{(\zeta)}_{\rm JK}\) for the dyonic loop gives
\begin{align}
Z^{(\zeta>0)}_{\rm JK}({\bm p}=1, \tilde{\bm p}=0, {\bm q}=1)&=\langle e^{{\rm i}  \oint (A^{(1d)}_t -{\rm i} \sigma^{(1d)}) dt } \rangle^{(\zeta>0)}_{\rm SQM}  \\
&=
e^{\frac{1}{2}(2\epsilon_+ +\sum_{f=1}^4 m_f)} \Bigl(\sum_{f=1}^4 e^{-m_f} \Bigr)
+e^{2\epsilon_+ +\frac{1}{2}(\sum_{f=1}^4 m_f)}  (e^{-a}+e^{a}) \nonumber \\
&\quad +\sum_{n=1}^{\infty} \tilde{c}_n({\bm m}, \epsilon_+)  e^{-n a}.
\label{eq:Dyonicexp}
\end{align}
We find that the first line of \eqref{eq:Dyonicexp} agrees with \(-Z^{(\zeta)}_{\rm extra}\) for the dyonic loop in \eqref{eq:dyonbub}. Thus, we can compute \(Z_{\rm mono}({\bm p}=1, \tilde{\bm p}=0, {\bm q}=1)\) by subtracting states from the VEV of the 1d Wilson loop.
Unlike in the case of the 't Hooft loop, for the dyonic loop we find that terms dependent on \(e^{\pm a}\) must also be removed. From a four-dimensional perspective, \(\langle e^{{\rm i}  \oint (A^{(1d)}_t -{\rm i} \sigma^{(1d)}) dt } \rangle^{(\zeta)}_{\rm SQM}\) is interpreted as the expectation value of the \(SU(2)\) fundamental Wilson loop in the (resolved) monopole bubbling background. 
Note that \(e^{2\epsilon_+ +\frac{1}{2}(\sum_{f=1}^4 m_f)}(e^{-a} + e^{a})\) has the same form as the VEV of a Wilson loop in the fundamental representation, up to the overall factor \(e^{2\epsilon_+ +\frac{1}{2}(\sum_{f=1}^4 m_f)}\). We therefore interpret this term as the contribution of a Wilson loop that does not couple to the monopole background, and hence it should be subtracted.

For higher magnetic charges, \(Z^{(\zeta)}_{\rm extra}\) in 't Hooft loops also contains states that depend on \(a\). Consequently, it is not straightforward to compute \(Z^{(\zeta)}_{\rm extra}\) precisely by using this expansion method. However, we expect that, in general, the states in \(Z^{(\zeta)}_{\rm JK}\) that are neutral under the 4d global gauge symmetry must be removed. We will see in Section \ref{sec:Koor} that this subtraction of neutral states works well for computing \(Z^{(\zeta)}_{\rm extra}\) for the lowest magnetic charge \({\bm p}={\bm e}_1\) in the \(Sp(N)\) gauge theory.

\section{Quantized Coulomb branch and spherical DAHA of $(C^{\vee}_1, C_1)$-type}
\label{sec:CCrank1}
In this section, we show that the VEVs of the loop operators computed in the previous section agree with the polynomial representation of the spherical DAHA of $(C^{\vee}_1, C_1)$-type.
Although it is known from various works  \cite{Oblomkov2004, Drukker:2009tz, Drukker:2009id, Alday:2009fs, Gaiotto:2010be, Ito:2011ea, Teschner:2013tqy, Cooke:2020KauffmanSkein, Hikami:2019jaw, Cirafici:2020qlf} that the deformation quantization of the Coulomb branch in the \(SU(2)\) gauge theory with four hypermultiplets should be identified with the spherical DAHA of $(C^{\vee}_1, C_1)$-type via the AGT relation, from the perspective of the quantization of flat \(SL(2, \mathbb{C})\) connections on a four-punctured sphere, we present a direct relation between BPS loop operators in the gauge theory and the polynomial representation of the spherical DAHA. See  also recent related works \cite{Allegretti:2024svn, Huang:2024mtw} for more details on this.
The new point of this section is to show that BPS loop operators in the gauge theory are directly related to the polynomial representation of the spherical DAHA.
On the other hand, since a picture in terms of a punctured Riemann surface is not yet known for the higher-rank cases, the relationship between the quantized Coulomb branch and the spherical DAHA studied in the next section is essentially new.


\subsection{Polynomial representation of the DAHA of $(C^{\vee}_1, C_1)$-type}
First, we introduce the necessary elements of the polynomial representation of the DAHA
to establish its identification with the BPS loop operators in the \(SU(2) \simeq Sp(1)\) gauge theory.
The DAHA of $(C^{\vee}_N, C_N)$-type, denoted by \(\mathcal{H}_{N}\) \cite{Sahi1999}, will be briefly discussed in Section \ref{sec:HooftKoor}. Here, we consider the rank-one case: \(N=1\). The DAHA of $(C^{\vee}_1, C_1)$-type, denoted by \(\mathcal{H}_{1}\), is the \(\mathbb{C}(q^{\frac{1}{2}},t^{\frac{1}{2}}_0,t^{\frac{1}{2}}_1,u^{\frac{1}{2}}_0,u^{\frac{1}{2}}_1)\)-algebra generated by 
\(T_0^{\pm1}, T_1^{\pm1}, T_0^{\vee \pm1}, T_1^{\vee \pm1}\) with 
the following relations \cite{NoumiStokman2004}:
\begin{align}
(T_0 -t_0^{\frac{1}{2}})(T_0+t^{-\frac{1}{2}}_0)&=0\,, \nonumber \\
(T_1 -t_1^{\frac{1}{2}})(T_1+t^{-\frac{1}{2}}_1)&=0\,, \nonumber \\
(T^{\vee}_0 -u_0^{\frac{1}{2}})(T^{\vee}_0+u_0^{-\frac{1}{2}})&=0\,,  \\
(T^{\vee}_1 -u_1^{\frac{1}{2}})(T^{\vee}_1+u_1^{-\frac{1}{2}})&=0\,, \nonumber \\
T^{\vee}_1 T_1 T^{\vee}_0 T_0&= q^{-\frac{1}{2}} \,. \nonumber
\end{align}
The spherical DAHA of $(C^{\vee}_1, C_1)$-type is defined by \(\mathrm{SH}_{1}:={\sf e} \mathcal{H}_{1} {\sf e}\), where
\({\sf e}\) is an idempotent given by
\begin{align}
{\sf e} &= \frac{1}{1+t_1}(1+t_1^{\frac{1}{2}}T_1),
\end{align}
which satisfies the relations
\begin{align}
{\sf e}^2={\sf e}, \quad 
{\sf e} T_1= T_1 {\sf e}=t_1^{\frac{1}{2}} {\sf e}.
\end{align}

The polynomial representation of \(\mathcal{H}_{1}\) is given by
\begin{align}
T_0 &\mapsto t^{\frac{1}{2}}_0 
+ t^{-\frac{1}{2}}_0 \frac{(1-u_0^{\frac{1}{2}} t_0^{\frac{1}{2}}  q^{\frac{1}{2}} x^{-1})
(1+u_0^{-\frac{1}{2}} t_0^{\frac{1}{2}} q^{\frac{1}{2}} x^{-1}) }{1- q x^{-2}} 
(s_0 -1) 
\,,\\
T_1&\mapsto t^{\frac{1}{2}}_1 
+ t^{-\frac{1}{2}}_1 \frac{(1-u_1^{\frac{1}{2}} t_1^{\frac{1}{2}}   x)
(1+u_1^{-\frac{1}{2}} t_1^{\frac{1}{2}}  x) }{1 -  x^2} (s_1 -1), \\
T^{\vee}_0 &\mapsto q^{-\frac{1}{2}} T^{-1}_0 x, \\
T^{\vee}_1 &\mapsto x^{-1} T^{-1}_1. 
\end{align}
Here, \(s_0\) and \(s_1\) are defined by \(s_1 f(x):=f(x^{-1})\) and \(s_0 f(x):=f(q x^{-1})\), respectively. The idempotent \({\sf e}\) projects onto 
the symmetric Laurent polynomial ring: \(\mathbb{C}[x] \to \mathbb{C}[x+x^{-1}]\), and \(\mathrm{SH}_{1}\) preserves \(\mathbb{C}[x+x^{-1}]\). The spherical DAHA is generated by \({\sf e}(T_1^{\vee} T_1 +(T_1^{\vee} T_1)^{-1}){\sf e}\), \({\sf e}(T_1T_0+(T_1T_0)^{-1}){\sf e}\), and \({\sf e}(T_1 T^{\vee}_0 + (T_1 T^{\vee}_0)^{-1}){\sf e}\). 
By a straightforward computation, we obtain the following expressions for the polynomial representation of these generators:
\begin{align}
{\sf e}(T_1^{\vee} T_1 +(T_1^{\vee} T_1)^{-1}){\sf e}& \mapsto x+x^{-1}, 
\label{eq:pol1} \\
{\sf e}(T_1T_0+(T_1T_0)^{-1}){\sf e}& \mapsto (t_0 t_1)^{-\frac{1}{2}} \Bigl( A_1(x) ( \hat{{\sf T}}-1)
+A_1(x^{-1}) (  \hat{{\sf T}}^{-1}-1) \Bigr)
\nonumber \\
& \quad+t_0^{\frac{1}{2}} t_1^{\frac{1}{2}}+t_0^{-\frac{1}{2}} t_1^{-\frac{1}{2}}, 
\label{eq:pol2}  \\
{\sf e}(T_1 T^{\vee}_0 + (T_1 T^{\vee}_0)^{-1} ){\sf e}& \mapsto 
(t_0 t_1)^{-\frac{1}{2}} \Bigl(q^{\frac{1}{2}} x A_1(x)(\hat{{\sf T}}-1)+ q^{\frac{1}{2}}  x^{-1} A_1(x^{-1}) (\hat{{\sf T}}^{-1}-1) 
\Bigr)\nonumber \\
&\quad+ t_1^{\frac{1}{2}} u_0^{\frac{1}{2}}  -t_1^{\frac{1}{2}} u_0^{-\frac{1}{2}}  +q^{\frac{1}{2}} ( t_0^{\frac{1}{2}}u_1^{\frac{1}{2}}
-t_0^{\frac{1}{2}} u_1^{-\frac{1}{2}}   )+  q^{\frac{1}{2}} t_0^{\frac{1}{2}}t_1^{\frac{1}{2}}
(x+x^{-1}).
\label{eq:pol3} 
\end{align}
Here, \(A_1(x)\) is defined as
\begin{align}
A_1(x) 
:= \frac{\left(1- q^{\frac{1}{2}} t_0^{\frac{1}{2}} u_0^{\frac{1}{2}} x   \right) \left(1+ q^{\frac{1}{2}} t_0^{\frac{1}{2}} u_0^{-\frac{1}{2}} x   \right)
\left(1- t_1^{\frac{1}{2}} u_1^{\frac{1}{2}} x \right) \left(1+ t_1^{\frac{1}{2}} u_1^{-\frac{1}{2}} x \right) }{(1-x^2) (1-q x^2)}.
\label{eq:AAA}
\end{align}
The operator \(\hat{{\sf T}}\) is a \(q\)-shift defined by \(\hat{{\sf T}} f(x):=f(q x)\).

\subsection{Deformation quantization of loop operators and quantized Coulomb branch}
Using the localization formula \eqref{eq:localization}, together with the monopole bubbling effects \eqref{eq:monothooft} and \eqref{eq:dyonbub}, 
the VEVs of the Wilson loop \(L_{(0,1)}\), the 't Hooft loop \(L_{(1,0)}\), and the dyonic loop \(L_{(1,1)}\) are given by
\begin{align}
\langle L_{(0,1)} \rangle &= e^{a}+e^{-a} \,, 
\label{eq:L01} \\
\langle L_{(1,0)} \rangle &=\left( e^{b} + e^{-b} \right) 
\left(\frac{\prod_{f=1}^4  \sh(\pm a-m_f) }{\sh(\pm2a)\sh(\pm2a+2\epsilon_+)} \right)^{\frac{1}{2}}
\nonumber \\
&\quad+\frac{\prod_{f=1}^4 \sh(a-m_f-\epsilon_+)}{\sh(2a) \sh(-2a+2\epsilon_+) }+\frac{\prod_{f=1}^4 \sh(-a-m_f-\epsilon_+)}{\sh(-2a) \sh(2a+2\epsilon_+) }
+\ch\Bigl(\sum_{f=1}^4 m_f +2 \epsilon_+\Bigr)\,,
\label{eq:L10} \\
\langle L_{(1,1)} \rangle&= ( e^{b+a}+e^{-b-a}  )\left(\frac{\prod_{f=1}^4  \sh(\pm a-m_f) }{\sh(\pm2a)\sh(\pm2a+2\epsilon_+)} \right)^{\frac{1}{2}}
\nonumber \\
&\quad+ e^{-a +\epsilon_+}\frac{\prod_{f=1}^4 \sh(a-m_f-\epsilon_+)}{\sh(2a) \sh(-2a+2\epsilon_+) }
+ e^{a +\epsilon_+} \frac{\prod_{f=1}^4 \sh(-a-m_f-\epsilon_+)}{\sh(-2a) \sh(2a+2\epsilon_+) }
\nonumber \\
&\qquad 
-e^{\frac{1}{2}(2\epsilon_+ +\sum_{f=1}^4 m_f)} \left(\sum_{f=1}^4 e^{-m_f} - e^{\epsilon_+-a}- e^{\epsilon_+ +a}\right)\,.
\label{eq:L11}
\end{align}

To establish the correspondence with the polynomial representation of the spherical DAHA of $(C_1^{\vee}, C_1)$-type, we introduce a variable \({\sf T}\) and rewrite the deformation quantization of loop operators:
\begin{align}
{\sf T}:=e^{b} \left( \frac{\prod_{f=1}^4 \sh (- a-m_f)}{\prod_{f=1}^4 \sh ( a-m_f)} \frac{\sh ( 2 a) \sh ( 2 a+ 2 \epsilon_+)}{\sh ( -2 a) \sh ( -2 a+ 2\epsilon_+)} \right)^{\frac{1}{2}}\,.
\end{align}
Since \(\hat{{\sf T}}\), which appears in \eqref{eq:pol2}, corresponds to the deformation quantization of \({\sf T}\), we use the same notation \(\hat{{\sf T}}\).
By applying the differential operator \(\exp(-\epsilon_+ \partial_{a} \partial_{b})\) to the localization formulas \eqref{eq:L01}, \eqref{eq:L10}, and \eqref{eq:L11},
we obtain the following expressions for the deformation quantization:
\begin{align}
\hat{L}_{(0,1)}&=e^{a} + e^{-a}, 
\label{eq:defloop1} \\
\hat{L}_{(1,0)}
&=
 \frac{\prod_{f=1}^4  \sh(a-m_f-\epsilon_+) }{\sh(2a-2\epsilon_+)\sh(2a)}( \hat{{\sf T}}-1)+  \frac{\prod_{f=1}^4  \sh(-a-m_f-\epsilon_+) }{\sh(-2a-2\epsilon_+)\sh(-2a)} (\hat{{\sf T}}^{-1}-1)
\nonumber \\
&\quad +\ch\Bigl(\sum_{f=1}^4 m_f +2 \epsilon_+\Bigr)\,, 
\label{eq:defloop2} \\
\hat{L}_{(1,1)}
&=
   e^{-a +\epsilon_+}\frac{\prod_{f=1}^4 \sh(a-m_f-\epsilon_+)}{\sh(2a) \sh(-2a+2\epsilon_+) }(\hat{{\sf T}}-1)
+ e^{a +\epsilon_+} \frac{\prod_{f=1}^4 \sh(-a-m_f-\epsilon_+)}{\sh(-2a) \sh(2a+2\epsilon_+) }(\hat{{\sf T}}^{-1}-1)
\nonumber \\
&
\quad -e^{\frac{1}{2}(2\epsilon_+ +\sum_{f=1}^4 m_f)} \left(\sum_{f=1}^4 e^{-m_f} - e^{\epsilon_+-a}- e^{\epsilon_+ +a} \right).
\label{eq:defloop3}
\end{align}
In the above equations, we define \(\hat{L}_{(p,q)}:=\widehat{\langle L_{(p,q)} \rangle }\) and write \(\hat{a}\) as \(a\) to simplify the expressions. Note that \(\hat{{\sf T}}\) acts on \(e^{-a}\) as \(\hat{{\sf T}}^{\pm}  e^{-a}=e^{\pm 2\epsilon_+} e^{-a} \hat{{\sf T}}^{\pm}\). If we identify the parameters in the gauge theory with those in the spherical DAHA of $(C_1^{\vee}, C_1)$-type as follows:
\begin{align}
e^{m_1+\epsilon_+}&=t_0^{\frac{1}{2}} u_0^{\frac{1}{2}} q^{\frac{1}{2}}\,, 
\,\,
e^{m_2+\epsilon_+}=-t_0^{\frac{1}{2}} u_0^{-\frac{1}{2}} q^{\frac{1}{2}}\,, 
\,\,
e^{m_3+\epsilon_+}=t_1^{\frac{1}{2}} u_1^{\frac{1}{2}}\,, \nonumber \\
e^{m_4+\epsilon_+}&=-t_1^{\frac{1}{2}} u_1^{-\frac{1}{2}}\,,  
\,\,
e^{-a}=x\,, 
\,\,
e^{2 \epsilon_+}=q\,,
\end{align}
then we find that the deformation quantization of loop operators \eqref{eq:defloop1}--\eqref{eq:defloop3} agree precisely with
the generators of the spherical DAHA of $(C_1^{\vee}, C_1)$-type \eqref{eq:pol1}--\eqref{eq:pol3}. Therefore, the quantized Coulomb branch, i.e., the algebra of loop operators 
generated by \eqref{eq:defloop1}--\eqref{eq:defloop3}, is identical to the spherical DAHA.

\section{Quantized Coulomb branch of $Sp(N)$ gauge theory and spherical DAHA of $(C^{\vee}_N, C_N)$-type}
\label{eq:CCNdaha}
We conjecture that the quantized Coulomb branch of the 4d $\mathcal{N}=2$ $Sp(N)$ gauge theory with four fundamental hypermultiplets and one hypermultiplet in the antisymmetric representation is isomorphic to the spherical DAHA of $(C^{\vee}_N, C_N)$-type. In this section, we provide evidence for this conjecture by studying Wilson loops, 
't Hooft loops, and elements of the polynomial representation of the DAHA.
 
\subsection{Polynomial representation of the DAHA of $(C^{\vee}_N, C_N)$-type}
\label{sec:HooftKoor}
The DAHA of $(C^{\vee}_N, C_N)$-type, denoted by $\mathcal{H}_N$, is an \(R:=\mathbb{C}(t^{\frac{1}{2}}_0, t^{\frac{1}{2}}_N, u^{\frac{1}{2}}_0, u^{\frac{1}{2}}_N, t^{\frac{1}{2}},q^{\frac{1}{2}})\)-algebra generated by $T_0,\cdots, T_N$ and variables $X_1, \cdots, X_N$ with the following relations \cite{Sahi1999}.
Here, \( \{ {T}_i \}_{i=0}^N \) are the generators of the affine Hecke algebra of ${C}_N$-type:
\begin{align}
(T_i-t^{\frac{1}{2}}_i)(T_i+t^{-\frac{1}{2}}_i)&=0 \quad \text{ for } i=0,\cdots, N, \quad t_1=\cdots =t_{N-1}=t, \nonumber \\
T_i T_j  &= T_j T_i \quad \text{ for } |i -j|>1, \quad i,j \neq 0,N\,, \nonumber \\
T_i T_{i+1} T_i  &= T_{i+1} T_i T_{i+1} \quad \text{ for } i=1,\cdots, N-2\,, \nonumber \\
T_i T_{i+1} T_i T_{i+1} &= T_{i+1} T_i T_{i+1} T_i \quad \text{ for } i=0, N-1\,.
\end{align}
It is known that the DAHA $\mathcal{H}_N$ has a PBW decomposition of the form
\[
R[X^{\pm}_1, \cdots, X^{\pm}_N] \otimes H_0 \otimes R[Y^{\pm}_1, \cdots, Y^{\pm}_N].
\]
Here, \( H_0 \) is the finite Hecke algebra of $C_N$-type generated by \( {T}_i \) for \( i=1, \cdots, N \), and \( Y_i \) are the Dunkl operators defined by
\begin{align}
{Y}_1&:= {T}_1 \cdots {T}_N {T}_{N-1} \cdots { T}_0 \,,\nonumber \\
{ Y}_2&:= { T}_2 \cdots { T}_N {T}_{N-1} \cdots { T}_0 {T}^{-1}_1\,, \nonumber \\
&\vdots \nonumber \\
{ Y}_N&:= { T}_N {T}_{N-1} \cdots { T}_0 { T}^{-1}_1 \cdots { T}^{-1}_{N-1} \,.
\end{align}
Note that the $Y_i$ commute with each other. The spherical DAHA is defined as ${\rm SH}_N:={\sf e} \mathcal{H}_N {\sf e}$, where ${\sf e}$ is an idempotent.
In the spherical DAHA, the Laurent polynomial ring in the \(X_i\) satisfies the relations
\begin{align} 
{\sf e} R[X^{\pm1}_1, \cdots, X^{\pm1}_N] {\sf e}&=R[X^{\pm1}_1, \cdots, X^{\pm1}_N]^{W_{Sp(N)}} 
\label{eq:WinvSH}
\,, \\
{\sf e} R[Y^{\pm1}_1, \cdots, Y^{\pm1}_N] {\sf e}&=R[Y^{\pm1}_1, \cdots, Y^{\pm1}_N]^{W_{Sp(N)}}\,.
\end{align} 
where \( W_{Sp(N)} \) is the Weyl group of type \( C_N \) (i.e., the Weyl group of the gauge group \( Sp(N) \)). Later, \eqref{eq:WinvSH} will be identified with the algebra of Wilson loops.

Next, we consider the polynomial representation. 
It was shown in \cite{Sahi1999} that Noumi's representation of the affine Hecke algebra \cite{Noumi_1995} extends to a representation of the DAHA:
\begin{align}
{ X}_i &\mapsto x_i\,, \\
{ T}_i &\mapsto t^{\frac{1}{2}}_i + t^{-\frac{1}{2}}_i \frac{1-t _i x_i x_{i+1}^{-1} }{1- x_i x_{i+1}^{-1}} 
(s_i -1) \quad (\text{ for } i=1,\cdots, N-1)\,, \\
{ T}_0 &\mapsto t^{\frac{1}{2}}_0 
+ t^{-\frac{1}{2}}_0 \frac{(1-u_0^{\frac{1}{2}} t_0^{\frac{1}{2}}  q^{\frac{1}{2}} x_{1}^{-1})
(1+u_0^{-\frac{1}{2}} t_0^{\frac{1}{2}} q^{\frac{1}{2}} x_{1}^{-1}) }{1- q x_{1}^{-2}} 
(s_0 -1) 
\,,\\
{T}_N &\mapsto t^{\frac{1}{2}}_N 
+ t^{-\frac{1}{2}}_N \frac{(1-u_N^{\frac{1}{2}} t_N^{\frac{1}{2}}   x_{N})
(1+u_N^{-\frac{1}{2}} t_N^{\frac{1}{2}}  x_{N}) }{1 -  x^2_{N}} (s_N -1)\,.
\end{align}
Here, the elements \( s_0,\cdots,s_N \) act on a function \( f(x_1,\cdots, x_N) \) as follows:
\begin{align}
s_i f(\cdots, x_i, x_{i+1}, \cdots)&=f(\cdots, x_{i+1},  x_{i}, \cdots) \quad \text{ for } i=1,\cdots, N-1\,,  \nonumber \\
s_0 f(x_1, x_2, \cdots)&=f(q x^{-1}_{1}, x_2, \cdots)\,, \\
s_N f( \cdots, x_{N-1}, x_N)&=f(\cdots, x_{N-1}, x^{-1}_N)\,. \nonumber
\end{align}
In particular, the representation of the degree-one elementary symmetric Laurent polynomial in \( Y_i+Y^{-1}_i \) is given by the 
Koornwinder operator \( V_1 \) up to an additive constant in \(R\) \cite{Noumi_1995}:
\begin{align}
{\sf e} \sum_{i=1}^N({ Y}_i+{ Y}^{-1}_i) {\sf e} \mapsto (t_0 t_N)^{-\frac{1}{2}} t^{1-N} \left( 
V_1 +(1+t_0t_N t^{N-1}  )  \frac{1- t^{N} }{1-t} \right)\,,
\end{align}
where the Koornwinder operator \( V_1 \) \cite{Koornwinder1992} is defined as
\begin{align}
V_1:=  \sum_{\varepsilon=\pm1} \sum_{i=1}^N   A_N(x^{\varepsilon}_i) \prod_{j= 1 \atop j \neq i }^N \frac{
  (1 -t x^{\varepsilon}_i  x_j  )
  (t x^{\varepsilon}_i -x_j ) }
{ 
 (1 - x^{\varepsilon}_i  x_j  )
  (x^{\varepsilon}_i -x_j )} ( \hat{\sf T}^{\varepsilon}_i -1) \,,
\end{align}
with
\begin{align}
A_N(x) 
:= \frac{\left(1- q^{\frac{1}{2}} t_0^{\frac{1}{2}} u_0^{\frac{1}{2}} x   \right) \left(1+ q^{\frac{1}{2}} t_0^{\frac{1}{2}} u_0^{-\frac{1}{2}} x   \right)
\left(1- t_N^{\frac{1}{2}} u_N^{\frac{1}{2}} x \right) \left(1+ t_N^{\frac{1}{2}} u_N^{-\frac{1}{2}} x \right) }{(1-x^2) (1-q x^2)}.
\end{align}

\subsection{'t Hooft loop \texorpdfstring{${L}_{({\bm e}_1,{\bm 0})}$}{L(e1,0)} and Koornwinder operator}
\label{sec:Koor}

To derive the localization formula for 't Hooft loops, we first recall the roots, the weights of the fundamental representation, and the weights of the second antisymmetric representation of \( Sp(N) \):

\begin{itemize}
\item The roots:
\begin{align}
\pm {\bm e}_{i}  \pm {\bm e}_{j} \quad (1 \le  i  <  j \le N), \quad  \pm 2 {\bm e}_i \quad (1 \le  i  \le  N)\,. 
\label{eq:rootssp}
\end{align}
\item The coroots: 
\begin{align}
  \pm ({\bm e}_{i} +{\bm e}_{j}) \quad (1 \le  i < j \le N), \quad \pm {\bm e}_{i} \quad (1 \le  i  \le  N)\,.
\end{align}
\item The weights of the fundamental representation:
\begin{align}
\pm {\bm e}_{i} \quad (1 \le  i  \le  N)\,.
\end{align}
\item The weights of the second antisymmetric representation:
\begin{align}
\pm {\bm e}_{i}  \pm {\bm e}_{j} \quad (1 \le  i  <  j \le N)\,.
\label{eq:antisymsp}
\end{align}
\end{itemize}
Here, \( {\bm e}_i := (0,\cdots, 0,\underset{i}{1},0,\cdots, 0) \) for \( i=1, \cdots, N \).

Using the one-loop determinant \eqref{eq:oneloopG} together with \eqref{eq:rootssp}--\eqref{eq:antisymsp}, the VEV of the 't Hooft loop \( \langle L_{({\bm e}_1,{\bm 0})} \rangle \) in the \( Sp(N) \) gauge theory is given by
\begin{align}
\langle L_{({\bm e}_1,{\bm 0})}\rangle &=\sum_{\varepsilon=\pm 1} \sum_{i=1}^N   e^{\varepsilon b_i} 
Z_{1\text{-loop}}({\bm p}=\varepsilon{\bm e}_i) +Z_{\rm mono}
({\bm p}={\bm e}_1,\tilde{\bm p}={\bm 0})\,,
\label{eq:vevSpN}
\end{align}
where
\begin{align}
Z_{1\text{-loop}}({\bm e}_i)&=Z_{1\text{-loop}}(-{\bm e}_i) \nonumber \\
&=\left( \frac{\prod_{f=1}^4
  {\rm sh} ( \pm a_i  -m_f  ) \prod_{j= 1 \atop j \neq i }^N
  {\rm sh} ( \pm a_i \pm a_j  -{m}_{\rm as}  ) }
{ {\rm sh} (  \pm 2 a_i  -2 \epsilon_+ ){\rm sh} ( \pm 2 a_i   )\prod_{j= 1 \atop j \neq i }^N {\rm sh} ( \pm  a_i \pm a_j -\epsilon_+   )}
\right)^{\frac{1}{2}} \,.
\label{eq:1loosp1}
\end{align}
Here, \( f(\pm x\pm y):=\prod_{s_1, s_2=\pm1}f(s_1 x+s_2 y) \). The parameters \( \{ m_f \}_{f=1}^4 \) and \( m_{\rm as} \) are the flavor fugacities for hypermultiplets in the four fundamental representations and the antisymmetric representation, respectively. The monopole bubbling effect
\( Z_{\rm mono} \) in \eqref{eq:vevSpN} is determined as follows.

For the rank-one case, brane configurations are available for the monopole bubbling effects and for their completion with extra D5-branes. However, 
for \( Sp(N) \) gauge theories, a systematic incorporation of hypermultiplets in the antisymmetric representation within both the naive and improved D-brane constructions remains unknown.\footnote{With an orientifold, naive brane configurations for the JK part of the monopole bubbling in \( Sp(N) \) and \( O(N) \) gauge theories with hypermultiplets in fundamental representations were constructed in \cite{Hayashi:2020ofu}.} Instead, we determine the monopole bubbling effects using a truncation of the instanton partition function, motivated by the Kronheimer correspondence. This correspondence establishes an isomorphism between the moduli space of the Bogomol'nyi equation with screened monopole charge and the moduli space of \( U(1) \)-invariant instantons on a single center Taub--NUT space. Consequently, \( Z^{(\zeta)}_{\rm JK} \) is obtained by truncating the 5d instanton partition function. The truncation is given as follows.

Let \( \sum_{\ell} e^{w_{\ell,{\rm p}}({\bm a}, \varepsilon_1,\varepsilon_2)} \) denote the \( \prod_{i} U(1)_{a_i} \times \prod_{l=1}^2 U(1)_{\varepsilon_l} \)-equivariant character of the tangent bundle of the ADHM moduli space for \( k \)-instantons, where \( w_{\ell,{\rm p}} \) denotes an equivariant weight at a torus fixed point \( {\rm p} \).\footnote{In general, \( w_{\ell,{\rm p}} \) depends on the choice of the FI-parameter \( \zeta \) (i.e., the stability condition) of the ADHM moduli space.}  We use $\varepsilon_{1,2}$ for the usual Nekrasov instanton $\Omega$-background parameters 
in order to distinguish them from the SQM chemical potentials $\epsilon_{1,2}$ introduced in Section \ref{sec:monobub}. 
To extract the \(U(1)_K\)-invariant sector associated with the Kronheimer correspondence, we denote by \(\epsilon_-\) the equivariant parameter for the \(U(1)_K\) isometry along the Taub--NUT circle fiber. Near the fixed point, the Taub--NUT space is locally described by \(\mathbb C^2\) with complex coordinates \((z_1,z_2)\), and the \(U(1)_K\) symmetry rotates them with opposite phases,
\begin{align}
(z_1,z_2)\longrightarrow (e^{-{\rm i}\alpha}z_1,e^{{\rm i}\alpha}z_2). 
\end{align}
Thus, the \(U(1)_K\) equivariant parameter enters the two Nekrasov equivariant weights with opposite signs. With the conventions used here, the embedding relevant for the \(U(1)_K\)-invariant instantons is implemented by the replacement
\begin{align}
 \varepsilon_{1}\mapsto\frac{\epsilon_{+}+\epsilon_-}{2}, \qquad \varepsilon_{2}\mapsto\frac{\epsilon_{+}-\epsilon_-}{2}, \qquad {\bm a}\mapsto{\bm a}+\tilde{\bm p}\,\epsilon_-\,. 
 \label{eq:U1replacement}
 \end{align}
Here \(\epsilon_+\) is the physical \(\Omega\)-background parameter of the loop-operator problem, whereas \( \epsilon_- \) is the equivariant parameter for the \(U(1)_K\) isometry of the Taub--NUT circle fiber. In particular, \(\epsilon_- \) should not be confused with the parameter \(m_*\) in Section \ref{sec:monobub}, which is the adjoint-hypermultiplet mass in the parent \(\mathcal N=2^*\) brane construction.
Then, the 5d Nekrasov partition function is given as a sum over the torus fixed points:
 \begin{align}
 Z^{(\zeta)}_{k\text{-inst}}({\bm a}, {\bm m}, \varepsilon_1, \varepsilon_2) = \sum_{{\rm p}: {\rm fixed}} \frac{f(w_{\ell,{\rm p}})}{\prod_{\ell} \sh(w_{\ell,{\rm p}})} \,.
 \end{align}
Here, \( f(w_{\ell,{\rm p}}) \) represents the hypermultiplet contribution, which depends on the representation of the hypermultiplets.
The claim of \cite{Ito:2011ea} is that \( Z^{(\zeta)}_{\rm JK} \) is obtained by considering the \( \epsilon_- \)-independent part of the instanton partition function:
 \begin{align}
  Z^{(\zeta)}_{\rm JK}({\bm p}, \tilde{\bm p},  {\bm q}={\bm 0}, {\bm a}, \epsilon_+ )= \sum_{{\rm p}: {\rm fixed}}  \frac{f(w_{\ell^{\prime},{\rm p}})}
  {\prod_{\ell^{\prime}} \sh(w_{\ell^{\prime},{\rm p}})} \,.
 \label{eq:U1inv}
 \end{align}
Here, \( w_{\ell^{\prime},{\rm p}} \) denote the equivariant weights that are independent of \( \epsilon_- \).

For example, let us explicitly perform the truncation of the Nekrasov formula for the one-instanton partition function in the \( SU(2) \) gauge theory with four hypermultiplets in the fundamental representation. The Nekrasov formula for the one-instanton partition function in five dimensions \cite{Nekrasov:2002qd, Nekrasov:2004vw, Hwang:2014uwa} is given by
\begin{align}
Z^{(\zeta>0)}_{1\text{-inst}}
&=\frac{1}{{\rm sh}(\varepsilon_1){\rm sh}(\varepsilon_2)} \left(\frac{\prod_{f=1}^4 \sh(a-m_f-\varepsilon_1-\varepsilon_2)}{\sh(2a) \sh(-2a+2\varepsilon_1+2\varepsilon_2) }+\frac{\prod_{f=1}^4 \sh(-a-m_f-\varepsilon_1-\varepsilon_2)}{\sh(-2a) \sh(2a+2\varepsilon_1+2 \varepsilon_2) } \right)\,, \\
Z^{(\zeta<0)}_{1\text{-inst}}
&=\frac{1}{{\rm sh}(\varepsilon_1){\rm sh}(\varepsilon_2)} \left(\frac{\prod_{f=1}^4 \sh(a-m_f+\varepsilon_1+\varepsilon_2)}{\sh(2a) \sh(-2a-2\varepsilon_1-2\varepsilon_2) }
+\frac{\prod_{f=1}^4 \sh(-a-m_f+\varepsilon_1+\varepsilon_2)}{\sh(-2a) \sh(2a-2\varepsilon_1-2 \varepsilon_2) } \right)\,.
\label{eq:1instSU2}
\end{align}
To extract the relevant part for monopole bubbling, we apply the truncation procedure using the replacement \eqref{eq:U1replacement} with \( \tilde{\bm p} = {\bm 0} \). As a result, the \( \epsilon_- \)-independent truncation is given by
\begin{align}
 &\text{The truncation of } Z^{(\zeta)}_{1\text{-inst}} 
\Big|_{\varepsilon_{1}  \mapsto \frac{ \epsilon_{+}+\epsilon_{-}}{2} \atop 
\varepsilon_{2}  \mapsto \frac{ \epsilon_{+}-\epsilon_{-}}{2}} 
\nonumber \\
 &= \left\{
\begin{aligned}
\frac{\prod_{f=1}^4 \sh(a-m_f-\epsilon_+)}{\sh(2a) \sh(-2a+2\epsilon_+) }+\frac{\prod_{f=1}^4 \sh(-a-m_f-\epsilon_+)}{\sh(-2a) \sh(2a+2\epsilon_+) } & \,\text{ for } \zeta > 0, \\
\frac{\prod_{f=1}^4 \sh(a-m_f+\epsilon_+)}{\sh(-2a) \sh(2a+2\epsilon_+) }+\frac{\prod_{f=1}^4 \sh(-a-m_f+\epsilon_+)}{\sh(2a) \sh(-2a+2\epsilon_+) } & \,\text{ for } \zeta < 0\,.
\end{aligned}
\right.
\label{eq:trun1in}
\end{align}
Therefore, \eqref{eq:trun1in} correctly reproduces \( Z^{(\zeta)}_{\rm JK} \) obtained from the brane construction \eqref{eq:JKp}.

In a similar manner, we apply the truncation of the instanton partition function to compute the JK part of the monopole bubbling effects in the \( Sp(N) \) gauge theory. 
In Appendix \ref{Appendix1}, we summarize the computation of JK parts obtained from one-instanton partition function. 
Using the replacement \eqref{eq:U1replacement} with \( \tilde{\bm p} = {\bm 0} \), the one-instanton partition function \( Z_{1\text{-inst}} \) in the \( Sp(N) \) gauge theory  
with four fundamental hypermultiplets and one hypermultiplet in the antisymmetric representation is given by
\begin{align}
Z_{1\text{-inst}}
&= \frac{1}{2{\rm sh}(-\epsilon_+ \pm \epsilon_-) {\rm sh}(m_{\rm as} \pm \epsilon_+)}  
  \nonumber \\
 & \quad \times  \Bigl(
  \prod_{f=1}^4 {\rm sh} ( m_f) \cdot  \prod_{i=1}^N \frac{  {\rm sh}(\pm a_i+ m_{\rm as} )}{ {\rm sh}(\pm a_i +\epsilon_+) }
 + 
 \prod_{f=1}^4 {\rm ch} ( m_f) \cdot \prod_{i=1}^N \frac{ {\rm ch}( \pm a_i+ m_{\rm as})}{ {\rm ch}(\pm a_i +\epsilon_+)} 
\Bigr)\,.
\label{eq:1instSp}
\end{align}
By applying the \( \epsilon_- \)-independent truncation to \eqref{eq:1instSp}, the JK part of the monopole bubbling effect is given by  
\begin{align}
Z_{\rm JK} 
({\bm e}_1,{\bm 0})
&= \frac{1}{2 {\rm sh}(m_{\rm as} \pm \epsilon_+)}  
  \Bigl(
  \prod_{f=1}^4 {\rm sh} ( m_f) \cdot  \prod_{i=1}^N \frac{  {\rm sh}(\pm a_i+ m_{\rm as} )}{ {\rm sh}(\pm a_i +\epsilon_+) }
 + 
 \prod_{f=1}^4 {\rm ch} ( m_f) \cdot \prod_{i=1}^N \frac{ {\rm ch}( \pm a_i+ m_{\rm as})}{ {\rm ch}(\pm a_i +\epsilon_+)} 
\Bigr)\,.
\label{eq:zmonosp}
\end{align}
It is worth noting that if the contribution from the hypermultiplet in the antisymmetric representation is removed from \eqref{eq:zmonosp}, the resulting expression should correspond to 
\( Z_{\rm JK} ({\bm e}_1,{\bm 0}) \) in the \( Sp(N) \) gauge theory with only four fundamental hypermultiplets. Indeed, \eqref{eq:zmonosp} without 
the antisymmetric hypermultiplet perfectly agrees with the JK part obtained via a brane construction with an orientifold in \cite{Hayashi:2020ofu}.

As discussed in Section \ref{sec:subt}, \( -Z_{\rm extra}({\bm p}=1, \tilde{\bm p}=0) \) represents the contribution of states decoupled from the \( SU(2) \simeq Sp(1) \) global gauge symmetry. For the higher-rank gauge group \( Sp(N) \), we conjecture that the extra term \( -Z_{\rm extra}({\bm p}={\bm e}_1, \tilde{\bm p}={\bm 0}) \) in the \( Sp(N) \) gauge theory is likewise given by states decoupled from the \( Sp(N) \) global gauge symmetry.
From the expansion of \( Z_{\rm JK} \), we obtain the extra term:
\begin{align}
Z_{\rm JK} ({\bm e}_1,{\bm 0})&=
\frac{e^{(N-1)m_{\rm as} -N \epsilon_+ -\frac{1}{2}\sum_{f=1}^4 m_f}e^{-N(m_{\rm as } -\epsilon_+)}}{(1-e^{-m_{\rm as } -\epsilon_+})(1-e^{-m_{\rm as } +\epsilon_+})} \Bigl( 1+ e^{\sum_{f=1}^4 m_f} +\sum_{1 \le k < l \le 4}e^{   m_k +m_l }  \Bigr)+O(e^{-a_i})\nonumber \\
&=: -Z_{\rm extra}({\bm e}_1,{\bm 0})+O(e^{-a_i})\,.
\label{eq:extrasp}
\end{align}

The monopole bubbling effect \( Z_{\rm mono} \), obtained via the truncation \eqref{eq:zmonosp} and the expansion \eqref{eq:extrasp}, is rewritten as
\begin{align}
Z_{\rm mono}({\bm e}_1,{\bm 0})&=Z_{\rm JK}({\bm e}_1,{\bm 0})+Z_{\rm extra} ({\bm e}_1,{\bm 0})
\label{eq:Spmono2p}
\\
&=-\sum_{\varepsilon=\pm1}\sum_{i=1}^N    \frac{\prod_{f=1}^4
  {\rm sh} ( \varepsilon a_i  -m_f  -\epsilon_+) 
  }
{ {\rm sh} (  - 2 \varepsilon a_i  +2  \epsilon_+ ){\rm sh} ( -2 \varepsilon a_i   )   } \prod_{j= 1 \atop j \neq i }^N \frac{{\rm sh} ( \varepsilon( -a_i \pm a_j ) -{m}_{\rm as}+\epsilon_+  )}{{\rm sh} (  \varepsilon(-a_i \pm a_j)   )}
\nonumber \\
&\quad 
+e^{\frac{N-1}{2}(m_{\rm as}-\epsilon_+)    }\left(\sum_{k=0}^{N-1} e^{- k (m_{\rm as}-\epsilon_+)} \right) {\rm ch}\Bigl(  (N-1) (m_{\rm as}-\epsilon_+)-2\epsilon_+ -\sum_{f=1}^4m_f \Bigr)
\,.
\label{eq:Spmono2}
\end{align}
We have verified the equality of \eqref{eq:Spmono2p} and \eqref{eq:Spmono2} for several values of \( N \) using \texttt{Mathematica}. 
As mentioned at the end of Section \ref{sec:imp}, the VEV of a loop operator should be invariant under the sign flip \( \epsilon_+ \to -\epsilon_+ \), which serves as a consistency check of our computation.
Indeed, we have verified that \eqref{eq:Spmono2} remains invariant under the sign flip for several values of \( N \).
Another consistency check is that \eqref{eq:Spmono2} for \( N=1 \) reproduces the monopole bubbling effect in the \( SU(2) \) gauge theory as computed via the improved brane construction \eqref{eq:monothooft}.

We define the deformation quantization of the VEV of the 't Hooft loop \eqref{eq:vevSpN} using the Weyl--Wigner transformation \eqref{eq:weylwig}. 
To establish a connection with the polynomial representation of the spherical DAHA, 
we express the deformation quantization \( \hat{L}_{({\bm e}_1, {\bm 0})} \) of the 't Hooft loop in terms of the operators \( {\sf T}_i \) for \( i=1,2, \dots, N \), defined by 
\begin{align}
{\sf T}_{i}&:=e^{b_i} \left( \frac{\prod_{f=1}^4 \sh (- a_i-m_f)}{\prod_{f=1}^4 \sh ( a_i-m_f)} 
\frac{\sh ( -2 a_i) \sh (- 2 a_i-2 \epsilon_+)}{\sh ( 2 a_i) \sh ( 2 a_i- 2\epsilon_+) 
}   \right.
\nonumber \\
&\qquad \qquad \times
 \left. \prod_{j= 1 \atop j \neq i }^N \frac{ \sh ( -  a_i  \pm a_j -\epsilon_+   ) \sh (  a_i   \pm a_j -{m}_{\rm as}   ) }
{  \sh (   a_i  \pm  a_j -\epsilon_+   ) \sh (  - a_i  \pm a_j -{m}_{\rm as}  ) }   \right)^{\frac{1}{2}}\,.
\end{align}
Applying the Weyl--Wigner transformation \eqref{eq:weylwig} along with the expression \eqref{eq:Spmono2} for the monopole bubbling effect, we obtain the following quantized expression for the 't Hooft loop:
\begin{align}
\hat{L}_{({\bm e}_1,{\bm 0})}&=\sum_{\varepsilon=\pm1}\sum_{i=1}^N   \frac{\prod_{f=1}^4
  {\rm sh} ( \varepsilon a_i  -{m}_f - \epsilon_+ ) \prod_{j= 1 \atop j \neq i }^N
  {\rm sh} ( \varepsilon(-a_i \pm a_j)  -m_{\rm as}+\epsilon_+ ) }
{ {\rm sh} (  - 2\varepsilon a_i   ){\rm sh} ( -2 \varepsilon a_i  + 2 \epsilon_+ )\prod_{j= 1 \atop j \neq i }^N {\rm sh} (  \varepsilon (-a_i \pm a_j )   ) }( \hat{\sf T}^{\varepsilon}_i -1)
\nonumber \\
&\quad +e^{\frac{N-1}{2} (m_{\rm as}-\epsilon_+)}\left(\sum_{k=0}^{N-1} e^{- k (m_{\rm as}-\epsilon_+)} \right) 
{\rm ch}\Bigl(  (1-N) m_{\rm as}+(N+1) \epsilon_+ +\sum_{f=1}^4m_f \Bigr)\,.
\label{eq:defsptHT}
\end{align}
Here, the operator \( \hat{\sf T}_i \) acts on \( e^{-a_j} \) as \( \hat{\sf T}_i e^{-a_j} =e^{-a_j + 2 \delta_{ij} \epsilon_+} \). For simplicity, we have omitted the hat notation for \( a_i \).

Next, we establish the correspondence between the deformation quantization of the 't Hooft loop and the spherical DAHA of \( (C^{\vee}_N, C_N) \)-type.
If we identify the variables in the gauge theory with those in the spherical DAHA as
\begin{align}
e^{m_1+\epsilon_+}&=t_0^{\frac{1}{2}} u_0^{\frac{1}{2}} q^{\frac{1}{2}}\,, 
\,\,
e^{m_2+\epsilon_+}=-t_0^{\frac{1}{2}} u_0^{-\frac{1}{2}} q^{\frac{1}{2}}\,, 
\,\,
e^{m_3+\epsilon_+}=t_N^{\frac{1}{2}} u_N^{\frac{1}{2}}\,, \nonumber \\
\,\,
e^{m_4+\epsilon_+}&=-t_N^{\frac{1}{2}} u_N^{-\frac{1}{2}}\,,  
\,\,
e^{-m_{\rm as}+\epsilon_+}=t\,, 
\,\,
e^{-a_i}=x_i\,, 
\,\,
e^{2 \epsilon_+}=q\,. 
\label{eq:idparameter}
\end{align}
Then, we find that the deformation quantization of the 't Hooft loop \eqref{eq:defsptHT} coincides with the polynomial representation of an element of the spherical DAHA, namely, the Koornwinder operator:
\begin{align}
{\sf e}  \sum_{i=1}^N({ Y}_i+{ Y}^{-1}_i) {\sf e} & \mapsto   (t_0 t_N)^{-\frac{1}{2}} t^{1-N} \left( 
V_1 +(1+t_0t_N t^{N-1}  )  \frac{1- t^{N} }{1-t} \right) =\hat{L}_{({\bm e}_1,{\bm 0})}.
\end{align}

\subsubsection*{Wilson Loops and the Spherical DAHA}

As mentioned at the end of Section \ref{sec:defqu}, the algebra of (gauge) Wilson loops is isomorphic to the $W_{Sp(N)}$-invariant Laurent polynomial ring, i.e., the symmetric Laurent polynomial ring of type $C_N$. In the path integral formalism, the quantities $\{e^{m_f}\}_{f=1}^4$, $e^{m_{\rm as}}$, and $e^{\epsilon_+}$ correspond to Wilson loops of the flavor $U(1)$ symmetries. 
Thus, using the identification of variables in \eqref{eq:idparameter} and the relation \eqref{eq:WinvSH}, the algebra of flavor and gauge Wilson loops
 can be identified with a subalgebra of the spherical DAHA as follows:
\begin{align}
\text{The algebra of flavor and gauge Wilson loops} \simeq R[x^{\pm1}_1, \cdots ,x^{\pm1}_N]^{W_{Sp(N)}} \subset {\rm SH}_N\,.
\end{align}

\subsection{'t Hooft Loop \texorpdfstring{${L}_{({\bm e}_1+\cdots+{\bm e}_k,{\bm 0})}$}{L(e1,0)} and the van Diejen Operator}

The Koornwinder operator belongs to a family of commuting difference operators known as the van Diejen operators $\{ V_k \}_{k=1}^N$ \cite{VanDiejen1995}:
\begin{align}
V_k =\sum_{J \subset \{1,\cdots, N\} \atop |J|=k } \sum_{j \in J} \sum_{\varepsilon_j=\pm1}
\sum_{l=1}^k (-1)^{l-1} \sum_{\emptyset \subsetneq J_1 \subsetneq \cdots \subsetneq J_{l} =J} \prod_{r=1}^{l} V_{\{\varepsilon\};J_r \backslash J_{r-1}; K_r} \left(\prod_{j \in J_1} \hat{\sf T}^{\varepsilon_j}_j-1\right)\,,
\end{align}
where $J_0 = \emptyset$, $K_r =\{1, \cdots, N \} \backslash J_r$, and 
\begin{align}
V_{\{\varepsilon\};J; K}
= \prod_{j \in J} A_N (x^{\varepsilon_j}_j) \cdot \prod_{i< j \atop i, j \in J} 
\frac{1- t x^{\varepsilon_i}_i x^{\varepsilon_j}_j}{1-x^{\varepsilon_i}_i x^{\varepsilon_j}_j } 
\frac{1- q t x^{\varepsilon_i}_i x^{\varepsilon_j}_j}{1-q x^{\varepsilon_i}_i x^{\varepsilon_j}_j } 
\cdot \prod_{i \in J, j \in K} 
\frac{1- t x^{\varepsilon_i}_i x_j}{1-  x^{\varepsilon_i}_i x_j } 
\frac{t x^{\varepsilon_i}_i - x_j}{ x^{\varepsilon_i}_i - x_j } \,.
\end{align}

In the polynomial representation of the spherical DAHA, the van Diejen operator $V_k$ of order $k$ arises from the elementary symmetric polynomial of degree $k$ in 
$Y_{i}+Y^{-1}_{i}$. We conjecture that a linear combination of $\hat{L}_{({\bm e}_1+{\bm e}_2+\cdots +{\bm e}_n , {\bm 0})}$ with $n=1,\cdots, k$ corresponds to the degree-$k$ elementary symmetric polynomial of 
$Y_{i}+Y^{-1}_{i}$, and hence to the van Diejen operator of order $k$\footnote{A similar story holds for the $\mathfrak{gl}(N)$-type case \cite{Okuda:2019emk}. In the 4d $\mathcal{N}=2^*$ $U(N)$ gauge theory, the deformation quantization of the localization formula $\hat{L}_{({\bm e}_{1}+\cdots+{\bm e}_k, {\bm 0})}$ is identified with the Macdonald operator of order $k$, which belongs to the polynomial representation of the $\mathfrak{gl}(N)$-type spherical DAHA. In this case, since monopole bubbling is absent for ${\bm p}={\bm e}_{1}+\cdots+{\bm e}_k$ in $U(N)$, the expectation value $\langle {L}_{({\bm e}_1+{\bm e}_2+\cdots +{\bm e}_k , {\bm 0})} \rangle $ is completely determined by the one-loop determinant.}.
However, since the extra terms for higher magnetic charges may depend on the Coulomb branch parameters $x_i$, it is difficult to determine these extra terms using the method in the previous section. We compute $\hat{L}_{({\bm e}_1+{\bm e}_2, {\bm 0})}$ up to $Z_{\rm extra}({\bm e}_1+{\bm e}_2, {\bm 0})$, and show that it partially agrees with $V_2$.

The localization formula for $\langle {L}_{({\bm e}_1+{\bm e}_2, {\bm 0})} \rangle$ is given by
\begin{align}
\langle L_{({\bm e}_1+{\bm e}_2,{\bm 0})}\rangle &=\sum_{1 \le i < j \le N}  (e^{ b_i +  b_j}
+ e^{- b_i - b_j})Z_{1\text{-loop}}({\bm e}_i+{\bm e}_j) 
+ \sum_{1 \le i < j \le N} (e^{ b_i -  b_j}+e^{-b_i +  b_j}) Z_{1\text{-loop}}({\bm e}_i-{\bm e}_j) \nonumber \\
&\quad + \sum_{1 \le i  \le N} (e^{ b_i}+e^{-b_i })  Z_{1\text{-loop}}({\bm e}_i) \, Z_{\rm mono} ({\bm e}_1+{\bm e}_2, {\bm e}_i)
+Z_{\rm mono}({\bm e}_1+{\bm e}_2, {\bm 0})\,.
\label{eq:vevSpNanti}
\end{align}
Here, the one-loop determinants are given by
{\small
\begin{align}
&Z_{1\text{-loop}}({\bm e}_i+{\bm e}_j) \nonumber \\
&= \left(\frac{\sh (\pm (a_i +a_j)-m_{\rm as} \pm \epsilon_+)}{\sh (\pm (a_i +a_j) )\sh (\pm (a_i +a_j) -2  \epsilon_+) } \prod_{k=i,j}\frac{\prod_{f=1}^4  \sh(\pm a_k-m_f) }{ \sh(\pm 2 a_{k}) \sh(\pm 2 a_{k}-2 \epsilon_+) } 
 \prod_{l = 1  \atop  l \neq i, j}^N  \frac{ \sh (\pm a_k  \pm a_l -m_{\rm as}) }{ \sh (\pm a_k \pm a_l -\epsilon_+) } \right)^{\frac{1}{2}}\,, \\
&Z_{1\text{-loop}}({\bm e}_i-{\bm e}_j) \nonumber \\
&= \left(\frac{\sh (\pm (a_i -a_j)-m_{\rm as} \pm \epsilon_+)}{\sh (\pm (a_i -a_j) )\sh (\pm (a_i -a_j) -2  \epsilon_+) } \prod_{k=i,j}\frac{\prod_{f=1}^4  \sh(\pm a_k-m_f) }{ \sh(\pm 2 a_{k}) \sh(\pm 2 a_{k}-2 \epsilon_+) } 
 \prod_{l = 1  \atop  l \neq i, j}^N  \frac{ \sh (\pm a_k  \pm a_l -m_{\rm as}) }{ \sh (\pm a_k \pm a_l -\epsilon_+) } \right)^{\frac{1}{2}}\,.
\end{align}
}
The JK part for $\tilde{\bm p}={\bm e}_i$ is given by \eqref{eq:ZJK12i}.
Since $Z_{\rm JK}({\bm e}_1+{\bm e}_2,{\bm e}_i)$ has the same form as $Z_{\rm JK}({\bm e}_1,{\bm 0})$ in the $Sp(N-1)$ gauge theory, with the $i$-th Coulomb branch parameter $a_i$ omitted, we conclude that $Z_{\rm extra}({\bm e}_1+{\bm e}_2,{\bm e}_i)$ is identical to $Z_{\rm extra}({\bm e}_1,{\bm 0})$ in the $Sp(N-1)$ gauge theory.

To compare with the van Diejen operator, we rewrite $\hat{L}_{({\bm e}_1+{\bm e}_2, {\bm 0})}$ using the parameter identification \eqref{eq:idparameter}:
{\small
\begin{align}
\hat{L}_{({\bm e}_1+{\bm e}_2, {\bm 0})} 
&=t^{3-2N} (t_0 t_N)^{-1} \sum_{\varepsilon_1,\varepsilon_2=\pm1}
\sum_{1 \le i_1 < i_2 \le N} \Bigl( \prod_{l =1}^2 A_N(x^{\varepsilon_l}_{i_l}) \cdot \prod_{ k \neq i_1, i_2}
\frac{1-t x^{\varepsilon_l}_{i_l} x_k}{1-x^{\varepsilon_l}_{i_l} x_k }\frac{x^{\varepsilon_l}_{i_l}-t  x_k}{x^{\varepsilon_l}_{i_l}- x_k } \Bigr) \nonumber \\
&  \times 
\Bigl\{ \frac{1-t x^{\varepsilon_1}_{i_1} x^{\varepsilon_2}_{i_2}}{1- x^{\varepsilon_1}_{i_1} x^{\varepsilon_2}_{i_2} }
\frac{1-q t x^{\varepsilon_1}_{i_1} x^{\varepsilon_2}_{i_2}}{1-q x^{\varepsilon_1}_{i_1} x^{\varepsilon_2}_{i_2} } \hat{{\sf T}}^{\varepsilon_1}_{i_1}\hat{{\sf T}}^{\varepsilon_2}_{i_2} 
+ \frac{1-t x^{\varepsilon_1}_{i_1} x_{i_2}}{1-x^{\varepsilon_1}_{i_1} x_{i_2} } \frac{x^{\varepsilon_1}_{i_1}-t  x_{i_2}}{x^{\varepsilon_1}_{i_1}- x_{i_2} } \hat{{\sf T}}^{\varepsilon_{1}}_{i_1}
+ \frac{1-t x^{\varepsilon_2}_{i_2} x_{i_1}}{1-x^{\varepsilon_2}_{i_2} x_{i_1} } \frac{x^{\varepsilon_2}_{i_2}-t  x_{i_1}}{x^{\varepsilon_2}_{i_2}- x_{i_1} }\hat{{\sf T}}^{\varepsilon_{2}}_{i_2} 
\Bigr\}
 \nonumber \\
&
- (q t_0 t_N)^{-\frac{1}{2}}   \frac{1-t^{N-1}}{1-t} (qt^{2-N}+ t_0 t_N) (\hat{L}_{({\bm e}_1, {\bm 0})}-
Z_{\rm mono}({\bm e}_1, {\bm 0})) 
   +Z_{\rm mono}({\bm e}_1+{\bm e}_2, {\bm 0}).
\label{eq:2ndL}
\end{align}
}
On the other hand, the van Diejen operator is given by
{\small
\begin{align}
V_2
&= \sum_{\varepsilon_1,\varepsilon_2=\pm1}
\sum_{1 \le i_1 < i_2 \le N} \Bigl( \prod_{l =1}^2 A_N(x^{\varepsilon_l}_{i_l}) \cdot \prod_{ k \neq i_1, i_2}
\frac{1-t x^{\varepsilon_l}_{i_l} x_k}{1-x^{\varepsilon_l}_{i_l} x_k }\frac{x^{\varepsilon_l}_{i_l}-t  x_k}{x^{\varepsilon_l}_{i_l}- x_k } \Bigr) 
\nonumber \\
& \times
\Bigl\{ \frac{1-t x^{\varepsilon_1}_{i_1} x^{\varepsilon_2}_{i_2}}{1- x^{\varepsilon_1}_{i_1} x^{\varepsilon_2}_{i_2} }
\frac{1-q t x^{\varepsilon_1}_{i_1} x^{\varepsilon_2}_{i_2}}{1-q x^{\varepsilon_1}_{i_1} x^{\varepsilon_2}_{i_2} }
( \hat{{\sf T}}^{\varepsilon_1}_{i_1}\hat{{\sf T}}^{\varepsilon_2}_{i_2} -1)
\nonumber \\
&  \qquad
- \frac{1-t x^{\varepsilon_1}_{i_1} x_{i_2}}{1-x^{\varepsilon_1}_{i_1} x_{i_2} } \frac{x^{\varepsilon_1}_{i_1}-t  x_{i_2}}{x^{\varepsilon_1}_{i_1}- x_{i_2} }
( \hat{{\sf T}}^{\varepsilon_{1}}_{i_1}-1)
- \frac{1-t x^{\varepsilon_2}_{i_2} x_{i_1}}{1-x^{\varepsilon_2}_{i_2} x_{i_1} } \frac{x^{\varepsilon_2}_{i_2}-t  x_{i_1}}{x^{\varepsilon_2}_{i_2}- x_{i_1} }
(\hat{{\sf T}}^{\varepsilon_{2}}_{i_2} -1)
\Bigr\}.
\label{eq:2ndVDop}
\end{align}
}
Comparing \eqref{eq:2ndL} with \eqref{eq:2ndVDop}, we find that, up to the overall normalization factor $t^{3-2N}(t_0t_N)^{-1}$, the coefficients of the shift-operator terms $\hat{{\sf T}}^{\varepsilon_1}_{i_1}\hat{{\sf T}}^{\varepsilon_2}_{i_2}$ and $\hat{{\sf T}}^{\varepsilon}_{i}$ agree. Therefore, the \(q\)-difference operator part of a suitable linear combination of \(\hat{L}_{({\bm e}_1+{\bm e}_2, {\bm 0})}\) and \(\hat{L}_{({\bm e}_1, {\bm 0})}\) coincides with that of the second van Diejen operator. However, since the \(x_i\)-dependent part of \(Z_{\rm extra}\) is still unknown, we cannot yet determine the terms without shift operators completely. We leave the full operator-level identification for future work.


\section{Summary and future directions}
\label{sec:summary}
In this paper, we have studied the relation between the quantized Coulomb branch of the $Sp(N)$ gauge theory and the spherical DAHA of $(C^{\vee}_N, C_N)$-type. For $N=1$, we showed that the generators of the algebra of loop operators agree with the polynomial representation of the spherical DAHA. For $N \ge 2$, we studied the 't Hooft loop $\hat{L}_{({\bm e}_1, {\bm 0})}$ and showed that its deformation quantization gives the Koornwinder operator appearing in the polynomial representation of the spherical DAHA. We also showed that the algebra of flavor and gauge Wilson loops is identified with the $C_N$-type symmetric Laurent polynomial ring in the polynomial representation of the spherical DAHA.

An important remaining problem is to determine $Z_{\rm extra}$ in the ${Sp}(N)$ gauge theory, in particular its dependence on the Coulomb branch parameters $a_i$. Solving this problem is necessary in order to establish the correspondence for more general loop operators and, in particular, to clarify the relation to higher van Diejen operators. One possible approach is to extend the brane setup for the ${Sp}(N)$ gauge theory in \cite{Hayashi:2019rpw} so as to include antisymmetric matter, and then complete the setup by adding extra D5-branes. It would be interesting to understand whether such a construction gives a systematic derivation of the extra terms, in parallel with the rank-one case.

We performed our analysis in four dimensions, but similar calculations can be carried out in three dimensions by replacing ${\sh}(x)$ with $x$ in the one-loop determinants and monopole bubbling effects \cite{Okuda:2019emk}. On the DAHA side, this corresponds to the rational degeneration. Therefore, the 3d ${Sp}(N)$ gauge theory with four fundamental hypermultiplets and one hypermultiplet in the antisymmetric representation is expected to be related to the rational spherical DAHA of $(C^{\vee}_N, C_N)$-type. On the other hand, performing a similar analysis in five dimensions on $T^2 \times \mathbb{R}^3$ should lead to elliptic difference operators, since the KK modes along $T^2$ deform the one-loop determinants and monopole bubbling contributions into Jacobi theta functions \cite{Yoshida:2021wot}. From this viewpoint, it would be interesting to study the relation between the deformation quantization of 't Hooft surface operators in \cite{Yoshida:2021wot} and the elliptic lift of van Diejen operators.

\section*{Acknowledgements}
The author would like to thank Saburo Kakei, Yoshiyuki Kimura, Hiraku Nakajima, Satoshi Nawata, and Shintaro Yanagida for helpful discussions. He also expresses his gratitude to Kohei Yamaguchi for  correspondence. This work was supported by the JSPS Grant-in-Aid for Scientific Research (Grant No. 21K03382).

\appendix
\label{appendix1}

\section{$Z_{\rm JK}$ in $Sp(N)$ gauge theory from instanton partition functions}
\label{Appendix1}
In Section \ref{sec:Koor}, we explained that $Z_{\rm JK}$ in $SU(2)$ gauge theory is obtained by truncating instanton partition functions. Here, we compute $Z_{\rm JK}({\bm e}_1, {\bm 0})$ and  $Z_{\rm JK}({\bm e}_1+{\bm e}_2, {\bm e}_i)$ by truncating instanton partition functions in $Sp(N)$ gauge theory. 

First, we summarize the localization formula for instanton partition functions in $Sp(N)$ gauge theory 
with $N_F$ hypermultiplets in the vector representation and a hypermultiplet in the antisymmetric representation \cite{Kim:2012gu, Hwang:2014uwa}.
The $k$-instanton partition function $Z_k$ of the 5d $\mathcal{N}=1$ $Sp(N)$ gauge theory is given by the Witten index of the $O(k)=O(k)_+ \sqcup O(k)_-$ gauged SQM:
\begin{align}
Z_k = \frac{1}{2} (Z^+_k +Z^{-}_k)\,.
\end{align}
Here, $Z^+_k$ (resp. $Z^-_k$) corresponds to the $O(k)_+$ (resp. $O(k)_-$) sector. 
From SUSY localization, $Z^{+}_k$ and $Z^-_k$ are given by the following JK residues: 
\begin{align}
Z^{\pm}_k = \frac{1}{|W_{O(k)_{\pm}}|} \oint_{{\rm JK(\zeta)}} \prod_{I} d u_I  
Z^{\pm}_{\rm vec} 
Z^{\pm}_{\rm fund} Z^{\pm}_{\rm anti}\,.
\end{align}
Here, $|W_{O(k)_{\pm}}|$ denotes the order of the Weyl group.
$Z^{\pm}_{\rm vec}$, $Z^{\pm}_{\rm fund}$, and $Z^{\pm}_{\rm anti}$ represent the contributions from the 5d $\mathcal{N}=1$ vector multiplet, hypermultiplets in the vector representation, and the hypermultiplet in the antisymmetric representation, respectively.
It is convenient to express the instanton number $k$ as $k= 2 n + \chi$ with $\chi =0,1$.
The order of the Weyl group $|W_{O(k)_{\pm}}|$ is given by
\begin{align}
|W_{O(k)_{+}}|=
2^{n-1+\chi} n !, \quad 
|W_{O(k)_{-}}|=
2^{n-1+\chi} (n-1+\chi) !\,.
\end{align}
$Z^{\pm}_{\rm vec}$, $Z^{\pm}_{\rm fund}$, and $Z^{\pm}_{\rm anti}$ are given as follows.

For $O(k)_+$, the contribution of the vector multiplet is given by
{\small{
 \begin{align}
Z^+_{\rm vec}&=\prod_{1 \le I < J \le n} \sh(\pm u_I \pm u_J) \cdot \Bigl( \prod_{I=1}^n \sh(\pm u_I) \Bigr)^{\chi}
\Bigl(\frac{1}{\sh(\pm \epsilon_- -  \epsilon_+)\prod_{i=1}^N \sh(\pm a_i -  \epsilon_+)  } \cdot
 \prod_{I=1}^n \frac{\sh(\pm u_I-2 \epsilon_+) }{\sh(\pm u_I \pm  \epsilon_--\epsilon_+)}\Bigr)^{\chi}\nonumber \\
&\times
\prod_{I=1}^n \frac{\sh(2\epsilon_+)}
{\sh(\pm \epsilon_- -  \epsilon_+)\sh(\pm 2 u_I\pm \epsilon_- -  \epsilon_+) 
\prod_{i=1}^N \sh(\pm u_I \pm a_i -\epsilon_+) } \prod_{1 \le I <  J \le n} \frac{\sh(\pm u_I \pm u_J - 2\epsilon_+)}{\sh(\pm u_I \pm u_J \pm \epsilon_-- \epsilon_+)}\,.
\end{align}
}}
For $O(k)_-$ with $k=2n+1$, the contribution of the vector multiplet is given by
{\small{
 \begin{align}
Z^-_{\rm vec}&=\prod_{1 \le I < J \le n} \sh(\pm u_I \pm u_J) \cdot  \prod_{I=1}^n \sh(\pm u_I) 
\frac{1}{\sh(\pm \epsilon_- -  \epsilon_+)\prod_{i=1}^N \ch(\pm a_i -  \epsilon_+)  } \cdot
 \prod_{I=1}^n \frac{\ch(\pm u_I-2 \epsilon_+) }{\ch(\pm u_I \pm  \epsilon_--\epsilon_+)}\nonumber \\
&\times
\prod_{I=1}^n \frac{\sh(2\epsilon_+)}
{\sh(\pm \epsilon_- -  \epsilon_+)\sh(\pm 2 u_I\pm \epsilon_- -  \epsilon_+) 
\prod_{i=1}^N \sh(\pm u_I \pm a_i -\epsilon_+) } \prod_{1 \le I <  J \le n} \frac{\sh(\pm u_I \pm u_J - 2\epsilon_+)}{\sh(\pm u_I \pm u_J \pm \epsilon_-- \epsilon_+)}\,.
\end{align}
}}
For $O(k)_-$ with $k=2n$, the contribution of the vector multiplet is given by
{\small{
 \begin{align}
Z^-_{\rm vec}&=\prod_{1 \le I < J \le n-1} \sh(\pm u_I \pm u_J) \cdot  \prod_{I=1}^{n-1} \sh(\pm u_I) \nonumber \\
& \times
\frac{\ch(2\epsilon_+)}{\sh(\pm \epsilon_- -  \epsilon_+) \sh(\pm 2 \epsilon_- - 2 \epsilon_+)\prod_{i=1}^N \ch(\pm 2 a_i -2  \epsilon_+)  } \cdot
 \prod_{I=1}^{n-1} \frac{\sh(\pm 2 u_I-4 \epsilon_+) }{\sh(\pm 2 u_I \pm 2 \epsilon_-- 2\epsilon_+)}\nonumber \\
&\times
\prod_{I=1}^{n-1} \frac{\sh(2\epsilon_+)}
{\sh(\pm \epsilon_- -  \epsilon_+)\sh(\pm 2 u_I\pm \epsilon_- -  \epsilon_+) 
\prod_{i=1}^N \sh(\pm u_I \pm a_i -\epsilon_+) } \cdot \prod_{1 \le I <  J \le n-1} \frac{\sh(\pm u_I \pm u_J - 2\epsilon_+)}{\sh(\pm u_I \pm u_J \pm \epsilon_-- \epsilon_+)}\,.
\end{align}
}}

For $O(k)_+$, the contribution of the hypermultiplet in the antisymmetric representation is given by
{\small{
 \begin{align}
Z^+_{\rm anti}&=\left( \frac{\prod_{i=1}^N \sh( m_{\rm as} \pm a_i) }{\sh( m_{\rm as} \pm \epsilon_+)} 
\prod_{I=1}^n  \frac{\sh(\pm u_I \pm m_{\rm as} - \epsilon_- ) }{\sh( \pm u_I \pm m_{\rm as} - \epsilon_+)} 
\right)^{\chi}
 \prod_{I=1}^n\frac{\sh(\pm m_{\rm as} -\epsilon_-)
 \prod_{i=1}^N \sh(\pm u_I \pm a_i-m_{\rm as})}{\sh(\pm m_{\rm as} -\epsilon_+) \sh(\pm 2 u_{I} \pm m_{\rm as} -\epsilon_+) } \nonumber \\
& \times \prod_{1 \le I < J \le n} \frac{\sh(\pm u_I \pm u_J \pm m_{\rm as}-\epsilon_-)}{\sh(\pm u_I \pm u_J \pm m_{\rm as}-\epsilon_+)}\,.
\end{align}
}}
For $O(k)_-$ with $k=2n+1$, the contribution of the hypermultiplet in the antisymmetric representation is given by
{\small{
 \begin{align}
Z^-_{\rm anti}&= \frac{\prod_{i=1}^N \ch( m_{\rm as} \pm a_i) }{\sh( m_{\rm as} \pm \epsilon_+)} 
\prod_{I=1}^n  \frac{\ch(\pm u_I \pm m_{\rm as} - \epsilon_- ) }{\ch( \pm u_I \pm m_{\rm as} - \epsilon_+)} 
\frac{\sh(\pm m_{\rm as} -\epsilon_-) \prod_{i=1}^N \sh(\pm u_I \pm a_i-m_{\rm as})}{\sh(\pm m_{\rm as} -\epsilon_+) \sh(\pm 2 u_{I} \pm m_{\rm as} -\epsilon_+) } \nonumber \\
& \times \prod_{1 \le I < J \le n} \frac{\sh(\pm u_I \pm u_J \pm m_{\rm as}-\epsilon_-)}{\sh(\pm u_I \pm u_J \pm m_{\rm as}-\epsilon_+)}\,.
\end{align}
}}
For $O(k)_-$ with $k=2n$, the contribution of the hypermultiplet in the antisymmetric representation is given by
{\small{
\begin{align}
Z^-_{\rm anti}
&=
\frac{
\ch(\pm m_{\rm as}-\epsilon_-)
\prod_{i=1}^N \sh(2m_{\rm as}\pm 2a_i)
}{
\sh(m_{\rm as}\pm\epsilon_+)
\sh(2m_{\rm as}\pm 2\epsilon_+)
}
\nonumber\\
&\quad\times
\prod_{I=1}^{n-1}
\frac{
\sh(\pm 2u_I\pm 2m_{\rm as}-2\epsilon_-)
}{
\sh(\pm 2u_I\pm 2m_{\rm as}-2\epsilon_+)
}
\frac{
\sh(\pm m_{\rm as}-\epsilon_-)
\prod_{i=1}^N
\sh(\pm u_I\pm a_i-m_{\rm as})
}{
\sh(\pm m_{\rm as}-\epsilon_+)
\sh(\pm 2u_I\pm m_{\rm as}-\epsilon_+)
}
\nonumber\\
&\quad\times
\prod_{1\le I<J\le n-1}
\frac{
\sh(\pm u_I\pm u_J\pm m_{\rm as}-\epsilon_-)
}{
\sh(\pm u_I\pm u_J\pm m_{\rm as}-\epsilon_+)
}\,.
\end{align}
}}

For $O(k)_+$, the contribution of $N_F$ fundamental hypermultiplets is given by
 \begin{align}
Z^+_{\rm fund}&= \prod_{f=1}^{N_F}\left(  \Bigl( \sh(m_f) \Bigr)^{\chi} \prod_{I=1}^n \sh(\pm u_I +m_f) \right)\,.
\end{align}
For $O(k)_-$ with $k=2n+1$, the contribution of $N_F$ fundamental hypermultiplets is given by
 \begin{align}
Z^-_{\rm fund}&= \prod_{f=1}^{N_F}\left(   \ch(m_f) \prod_{I=1}^n \sh(\pm u_I +m_f) \right)\,.
\end{align}
For $O(k)_-$ with $k=2n$, the contribution of $N_F$ fundamental hypermultiplets is given by
 \begin{align}
Z^-_{\rm fund}&= \prod_{f=1}^{N_F}\left(   \sh(2m_f) \prod_{I=1}^{n-1} \sh(\pm u_I +m_f) \right)\,.
\end{align}

Let us write down \( Z_{\rm JK} \) obtained from \( Z_{k\text{-inst}} \) with \( k=1 \) in the \( Sp(N) \) gauge theory with \( N_F \) fundamental hypermultiplets and one hypermultiplet in the antisymmetric representation. The 1-instanton partition function is given by
\begin{align}
Z_{1\text{-inst}}
&= \frac{1}{2{\rm sh}(\epsilon_+ \pm \epsilon_-) {\rm sh}(m_{\rm as} \pm \epsilon_+)}  
  \nonumber \\
 & \times  \Bigl(
  \prod_{f=1}^{N_F} {\rm sh} ( m_f) \cdot  \prod_{i=1}^N \frac{  {\rm sh}(\pm a_i+m_{\rm as} )}{ {\rm sh}(\pm a_i -\epsilon_+) }
 + 
 \prod_{f=1}^{N_F} {\rm ch} ( m_f) \cdot \prod_{i=1}^N \frac{ {\rm ch}( \pm a_i+ m_{\rm as})}{ {\rm ch}(\pm a_i -\epsilon_+)} 
\Bigr)\,.
\label{eq:app1ins}
\end{align}
Then, \( Z_{\rm JK}({\bm e}_1,{\bm 0}) \) is given by the \( \epsilon_- \)-independent part of \eqref{eq:app1ins}:
\begin{align}
Z_{\rm JK}({\bm e}_1,{\bm 0})&={\rm sh}(\epsilon_+ \pm \epsilon_-) Z_{1\text{-inst}} \nonumber \\
&= \frac{1}{2 {\rm sh}(m_{\rm as} \pm \epsilon_+)}  
  \Bigl(
  \prod_{f=1}^{N_F} {\rm sh} ( m_f) \cdot  \prod_{j=1}^N \frac{  {\rm sh}(\pm a_j+m_{\rm as} )}{ {\rm sh}(\pm a_j-\epsilon_+) }
 + 
 \prod_{f=1}^{N_F} {\rm ch} ( m_f) \cdot \prod_{j=1}^N \frac{ {\rm ch}( \pm a_j+ m_{\rm as})}{ {\rm ch}(\pm a_j-\epsilon_+)} 
\Bigr)\,.
\label{eq:ZJK10}
\end{align}
Next, we consider \( Z_{\rm JK}({\bm e}_1+{\bm e}_2,{\bm e}_i) \). From \eqref{eq:U1inv}, \( Z_{\rm JK}({\bm e}_1+{\bm e}_2,{\bm e}_i) \) is given by the \( \epsilon_- \)-independent part of the 1-instanton partition function \eqref{eq:app1ins} after 
the shift \( {\bm a} \mapsto {\bm a}+\epsilon_- {\bm e}_i \):
\begin{align}
Z_{\rm JK}({\bm e}_1+{\bm e}_2,{\bm e}_i)
= \frac{1}{2 {\rm sh}(m_{\rm as} \pm \epsilon_+)}  
  &\Bigl(
  \prod_{f=1}^{N_F} {\rm sh} ( m_f) \cdot  \prod_{j=1 \atop j \neq i}^N \frac{  {\rm sh}(\pm a_j+m_{\rm as} )}{ {\rm sh}(\pm a_j -\epsilon_+) }
 \nonumber \\ 
 &\qquad \qquad + 
 \prod_{f=1}^{N_F} {\rm ch} ( m_f) \cdot \prod_{j=1 \atop j\neq i}^N \frac{ {\rm ch}( \pm a_j+ m_{\rm as})}{ {\rm ch}(\pm a_j-\epsilon_+)} 
\Bigr)\,.
\label{eq:ZJK12i}
\end{align}

\bibliography{refs}

\end{document}